\documentclass[aps,pra,twocolumn,twoside,10pt]{revtex4-1}

\usepackage{graphicx,epstopdf,amsmath,amssymb,braket,color,epsfig,epstopdf,geometry}
\usepackage[colorlinks=true, citecolor=red, urlcolor=blue]{hyperref}
\usepackage[table]{xcolor}
\usepackage{times,array,multirow}

\begin{document}

\title{
Conclusive Identification of Quantum Channels via Monogamy of Quantum Correlations
} 
\author{Asutosh Kumar\(^{1,2}\), Sudipto Singha Roy\(^{1,2}\), Amit Kumar Pal\(^{1,2}\), R. Prabhu\(^{1,2,3}\), Aditi Sen(De)\(^{1,2}\), and 
Ujjwal Sen\(^{1,2}\)}
\affiliation{\(^1\)Harish-Chandra Research Institute, Chhatnag Road, Jhunsi, Allahabad 211019, India\\
\(^2\)Homi Bhabha National Institute, Training School Complex, Anushaktinagar, Mumbai 400094, India\\
\(^3\)Department of Physics, Indian Institute of Technology Patna, Bihta 801103, Bihar, India​}
\pacs{}

\begin{abstract}

We investigate the action of local and global noise on monogamy of quantum correlations, when 
monogamy scores are considered as observables, and three-qubit 
systems are subjected to global noise and various local noisy channels, namely, amplitude-damping, phase-damping, and depolarizing channels. 
We show that 
the dynamics of monogamy scores corresponding to negativity and quantum discord, 
in the case of generalized W states, as inputs to the noisy channels, can exhibit non-monotonic dynamics with respect to 
increasing noise parameter, which is in contrast 
to the monotonic decay of 
monogamy scores when generalized Greenberger-Horne-Zeilinger states are exposed to noise. We quantify the persistence of monogamy 
against noise via a characteristic value of the noise parameter, and show that depolarizing noise destroys monogamy of quantum correlation 
faster compared to other noisy channels. We 
demonstrate that the negativity monogamy score is more robust than the quantum discord monogamy score, 
when the noise is of the phase-damping type. 
We also investigate the variation of monogamy with increasing noise for 
arbitrary three-qubit pure states as inputs. Finally, depending on these results, we propose a two-step protocol, which can conclusively 
identify the type of noise applied to the quantum system, by using generalized Greenberger-Horne-Zeilinger and generalized W 
states as resource states. We discuss a possible generalization of the results to higher number of parties.
\end{abstract}

\maketitle

\section{Introduction}
\label{sec:intro}

An important characterization of a composite quantum system is by the correlations, both classical and quantum, between its constituting parts. 
Quantum information theory provides a collection of measures of quantum correlations \cite{ent_horodecki,celeri_2011,modi_rmp_2012}, 
which can broadly be categorized into two classes. One is the ``entanglement-separability'' class, 
encompassing various measures of quantum entanglement in both bipartite and multipartite domain \cite{ent_horodecki}. 
The other is the information-theoretic regime \cite{celeri_2011,modi_rmp_2012}, consisting of quantum correlations such as 
quantum discord \cite{disc_group}, and various ``discord-like'' measures \cite{wdef_group,qd_geom},  that 
quantify quantum correlations beyond entanglement. 
Both entanglement as well as information theoretic quantum correlation measures 
have been proposed to be resources for several quantum protocols 
\cite{densecode,communication,teleport,ASDUS,onewayQC,diqc,crypto,nl_w_ent,opt_det,dqci,op_int,qd_geom_use}, 
and have been observed successfully in the laboratory \cite{ent_exp,disc_expt}.
However, quantum correlations, especially entanglement, have been found to be fragile under decoherence \cite{decohere_rev}.
Naturally, due to their immense importance in quantum information processing tasks, 
investigating the behavior of quantum correlations under various kinds of environmental noise has been a topic 
of utmost importance in quantum information theory.

Most of the available literature that deals with decoherence of quantum correlations consider bipartite
quantum correlation measures due to their relative computational simplicity 
\cite{sd_group,sd_group2,disc_dyn_group,disc_dyn_group2,disc_dyn_group3,disc_freez_group,chanda_freeze,ent_freez}. 
It has been shown that the bipartite entanglement measures tend to decay rapidly with 
increasing noise, and vanish when a threshold noise level is crossed. This phenomena is known as ``entanglement
sudden death'', and has been studied extensively in the case of bipartite systems under different types of environments \cite{sd_group,sd_group2}. 
In stark contrast to this behavior, information theoretic measures, namely, quantum discord, quantum work 
deficit, and several geometric measures, have been found to undergo an asymptotic decay with increasing 
noise strength \cite{mazeiro_disc_dyn,disc_dyn_group,disc_dyn_group2,disc_dyn_group3}, 
indicating a higher robustness against noise than that of entanglement. It has also been shown that special two- as well as multiqubit 
mixed quantum states can be engineered for which ``discord-like'' quantum correlations may remain frozen over 
a finite range of noise strength \cite{disc_freez_group,chanda_freeze}, while the entanglement measures for those states exhibit no such 
property (cf. \cite{ent_freez}). Although behavior of bipartite quantum correlations under decoherence is a well-investigated topic, 
similar studies in the multipartite scenario \cite{multi_decohere} are limited due to the lack of computable measures of quantum 
correlations for mixed multipartite states.  
 
Recent developments on the monogamy relation of quantum correlations \cite{mono,mono_terhal,mono_ent,koashi,mono_disc_other} 
have provided an effective tool to investigate
the multipartite nature of quantumness present in a composite quantum system. Qualitatively, monogamy of a quantum 
correlation measure corresponding to a multipartite state is the property that allows a chosen party to share only 
limited amount of quantum correlation with all the other parties except one, to which it is highly quantum correlated. 
Interestingly, such monogamy constraints can be quantified via the ``monogamy score'' \cite{mono_score},  leading to multipartite quantum 
correlation measures that use
bipartite measures of quantum correlations, thereby reducing the difficulty in the computation of the 
measures for multipartite states. The monogamy property of quantum correlations 
has been shown to be important in several aspects in quantum mechanics and quantum information, 
like foundations of quantum mechanics \cite{mono_found}, 
quantum cryptography \cite{mono_crypto}, teleportation \cite{mono_teleport}, quantum dense coding \cite{mono_densecode}, 
quantum steering \cite{mono_steering}, many-body physics \cite{mono_manybody}, and black-hole information theory \cite{mono_blackhole}.
Experimental investigation of this property has also been initiated \cite{mono_expt}. 
Therefore, it has become important to investigate the behavior of the monogamy property of quantum correlations when the system is 
subjected to noisy environments. 

This paper has two different objectives that are complementary to each other. 
In one, we study the dynamics of monogamy of quantum correlations. As measures of quantum correlations, we 
use the monogamy scores of two bipartite quantum correlation measures, namely, the negativity \cite{neg_gr,neg_part}, a measure of bipartite 
entanglement, and quantum discord \cite{disc_group,total_corr}, a quantum correlation measure 
from the information-theoretic domain. 
We choose a global noise, and three local noisy 
channels, namely, the amplitude-damping (AD), the phase-damping (PD), and the depolarizing (DP) channels
as different models of environmental noise \cite{decohere_rev,mazeiro_disc_dyn,operator-sum,kraus,preskill,sudarshan,choi}. 
We demonstrate how the dynamics of monogamy, in the case of three-qubit systems, exhibit qualitatively different behavior depending on whether 
the input quantum state is chosen from the family of generalized Greenberger-Horne-Zeilinger (gGHZ) state \cite{ghzstate}, or the 
generalized W (gW) states \cite{zhgstate,W-vs-GHZ,dvc}, which are not equivalent under stochastic local operations and classical 
communication (SLOCC). More specifically, we show that 
monogamy scores of negativity as well as quantum discord exhibit a 
monotonic decay with respect to the corresponding noise parameter, when gGHZ state is subjected to these noise 
models, while there exist non-monotonic dynamics when the input 
state is the gW state.
We also investigate the trends of monogamy scores against noise, 
when arbitrary three-qubit pure states belonging to the two inequivalent SLOCC classes of three-qubit pure states, 
namely, the GHZ and the W classes \cite{dvc}, are 
chosen as inputs.  Moreover, we introduce a concept called the ``dynamics terminal'', which quantify the durability of quantum correlation
measures under decoherence, and 
show that it can distinguish between different quantum correlation measures as well as different types of noise. 
The study also reveals that for the PD channel, the
negativity monogamy score can exhibit a more robust behavior against noise strength than that observed for the monogamy score of quantum 
discord, which we call the ``discord monogamy score''.

Besides characterizing the dynamical features of quantum correlations under decoherence, it is also interesting to address 
the reverse question as to whether the modes of environmental noise can be identified by using the properties of quantum correlations. 
Although a few studies have been motivated by similar goal \cite{channel_identify}, the literature regarding this issue is extremely limited. 
While most of the studies have tried to distinguish different types of noise by the different dynamical behavior of 
different quantum correlations, concrete protocol to conclusively identify the type of noise to which the quantum state is 
exposed is yet to be introduced. As the second objective of this paper, we 
use the highly entangled gGHZ and any gW states as resources, and
design a two-step protocol involving the monogamy relations of negativity and quantum discord to conclusively distinguish 
the type of noise applied to the quantum state, where the noise models include a global noise, and several local channels, 
namely, AD, PD, and DP channels. 

The paper is organized as follows. 
In Sec. \ref{sec:monogamy_decohere}, we discuss the dynamical behavior of the negativity and discord monogamy scores, 
when the gGHZ and  the gW states are subjected to different types of noise. The behavior of monogamy against noise, when arbitrary 
three-qubit pure states are considered as input, is also studied in this section.  
In Sec. \ref{sec:discrimination-tool}, the two-step channel discrimination 
protocol with monogamy scores is presented. Sec. \ref{sec:summary} presents the concluding remarks. 

%%%%%%%%%%%%%%%%%%%%%%%%%%%%%%

\begin{figure}
 \includegraphics[scale=0.3]{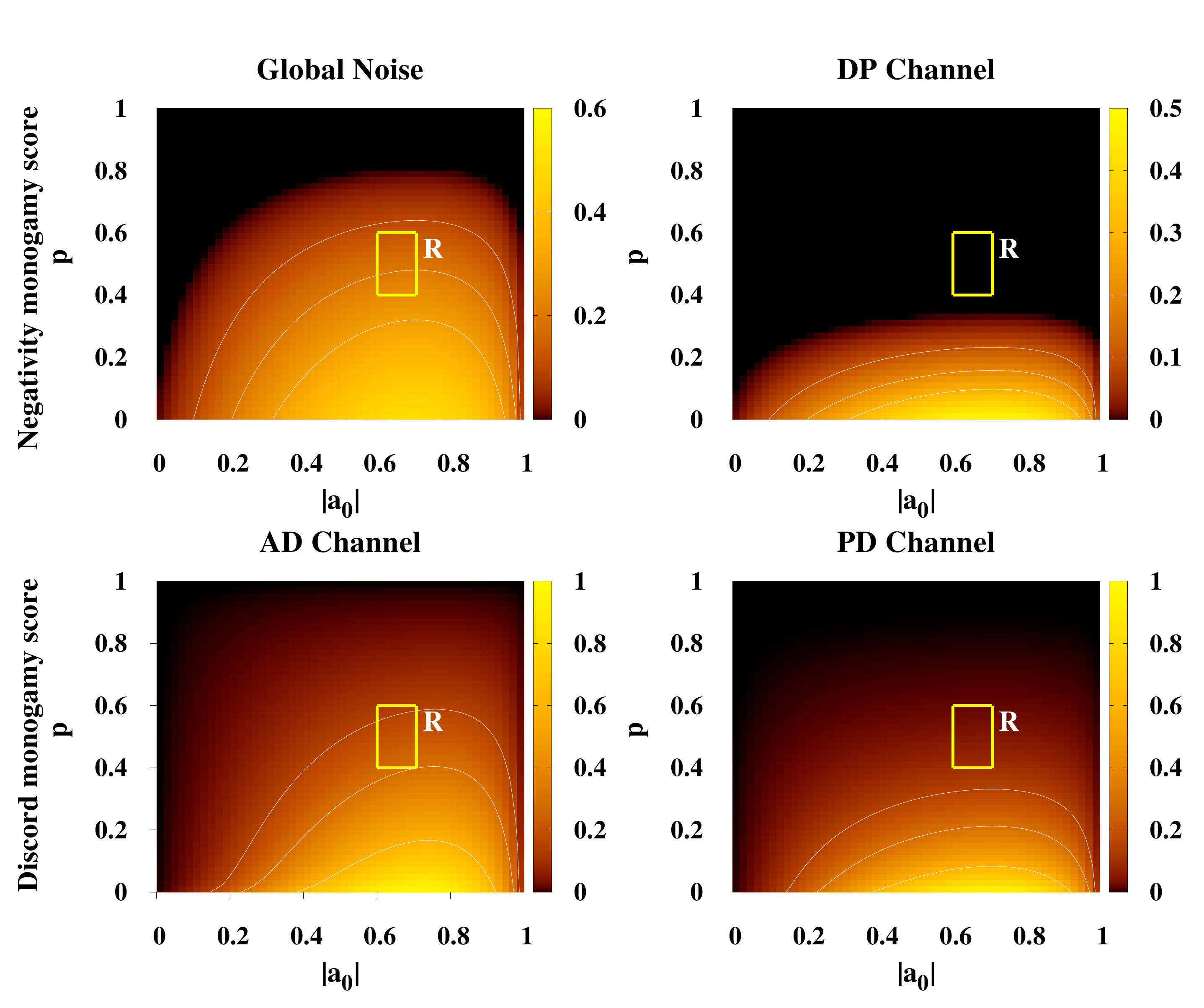}
 \caption{(Color online.) (Top panels) Variations of $\delta_{\mathcal{N}}$ as functions of 
 the noise parameter, $p$, and the state parameter, $|a_0|$, when gGHZ states are subjected to
 global noise (left) and DP channel (right). 
 (Bottom panels) Variations of $\delta_{\mathcal{D}}$ as functions of 
 the noise parameter, $p$, and the state parameter, $|a_0|$, when gGHZ states are subjected to
 AD channel (left) and PD channel (right).
 The absolute value of the other state parameter, $|a_1|$, 
 is determined by normalization. The solid lines in the plots are the contours obtained by joining the points 
 corresponding to a fixed value of either $\delta_{\mathcal{N}}$, or $\delta_{\mathcal{D}}$. In the case of $\delta_{\mathcal{N}}$, 
 the lines, from low to high values of $p$, correspond to $\delta_{\mathcal{N}}=0.3,0.2$, and $0.1$, while for $\delta_\mathcal{D}$,
 they represent the contours of $\delta_\mathcal{D}=0.6,0.3$, and $0.15$.  
 The regions marked by ``\textbf{R}'', and enclosed by the boxes are defined by the ranges $0.65\leq|a_0|\leq0.7071$, and
 $0.4\leq p\leq0.6$. The implications of these ranges of values are discussed in Sec. \ref{sec:discrimination-tool}.
 All the quantities plotted are dimensionless, except for $\delta_\mathcal{D}$, which is in bits.}
 \label{gghz}
\end{figure}

\section{Monogamy of quantum correlations under decoherence}
\label{sec:monogamy_decohere}

In this section, we investigate the behavior of monogamy scores corresponding to negativity and quantum discord for three-qubit
quantum states under the influence of global as well as local noise. Brief descriptions of the quantum correlation measures, namely, 
negativity and quantum discord, and monogamy of quantum correlations can be found in \cite{supp}. 
Discussions on different types of noise considered in this paper are provided in \cite{supp}.
Before considering arbitrary three-qubit pure states,
we examine the generalized GHZ (gGHZ), and the generalized W (gW) states as the input states to various 
types of noise.

\subsection{Generalized GHZ states}
\label{sec:gghz}
The generalized GHZ state, shared between three qubits, $1$, $2$, and $3$, reads as
\begin{equation}
|\Psi_3\rangle = a_0|000\rangle+a_1|111\rangle,
\label{eq:gGHZ}
\end{equation}
where $a_0$ and $a_1$ are the complex parameters satisfying $|a_0|^2+|a_1|^2=1$. 
In this paper, we consider qubit $1$ as the nodal observer while computing monogamy scores for negativity ($\delta_\mathcal{N}$),
and quantum discord ($\delta_{\mathcal{D}}$). 
Note that the monogamy scores for the gGHZ state, in the noiseless scenario, is 
always positive for all quantum correlation measures 
including negativity and quantum discord. This is due to the fact that the two-qubit reduced density matrix
$\rho_{12}=\rho_{13}=|a_0|^2|00\rangle\langle 00|+|a_1|^2|11\rangle\langle 11|$,
obtained from the gGHZ state, is a classically correlated two-qubit state having vanishing quantum correlations, 
while the state $|\Psi_3\rangle$ in the \(1:23\) bipartition always has a non-zero value of quantum 
correlation for $a_0,a_1\neq0$. Even for the noise parameter $p\neq 0$,   
$\delta_{\mathcal{N}}$ and $\delta_{D}$ are given by 
$\delta_{\mathcal{N}}=\mathcal{N}\big(\rho^{\mbox{\scriptsize gGHZ\normalsize}}_{1:23}\big)$ and 
$\delta_\mathcal{D}=\mathcal{D}\big(\rho^{\mbox{\scriptsize gGHZ\normalsize}}_{1:23}\big)$ respectively,
when the gGHZ state is subjected to the four types of noise considered in this paper (see \cite{supp}). 
Hence, both negativity and quantum discord are always monogamous in the present scenario, which can be applied to 
discriminate channels, as we shall see in Sec. \ref{sec:discrimination-tool}. 
Note that the entire discussion also holds for gGHZ states of an arbitrary number of parties
subjected to the different types of local and global noise considered here.

The above discussion helps one to determine analytical expressions for negativity monogamy score as a function of the noise parameter
in the case of different types of noise \cite{supp}.  
On the other hand, analytically determining the discord monogamy score, $\delta_\mathcal{D}$, for all the types of noise, 
is in general hard due to the optimization required to compute quantum discord for $\rho^{\mbox{\scriptsize gGHZ\normalsize}}$ 
in the $1:23$ split \cite{huang_disc_np}. So far, analytical determination of quantum discord has been possible only for very restricted class
of mixed states \cite{luo_pra_2008,disc_2q}. 
Hence, we employ numerical optimization over the real parameters $(\theta,\phi)$ of measurement involved in the definition of quantum discord.
The behavior of the monogamy scores corresponding to negativity and quantum discord for 
different types of noise are depicted in Fig. \ref{gghz}, where the top panels are for $\delta_{\mathcal{N}}$, and the bottom 
panels correspond to $\delta_{\mathcal{D}}$. For all the noise models considered in this paper, $\delta_{\mathcal{N}}$ and 
$\delta_{\mathcal{D}}$ monotonically decreases with increasing values of $p$ for a fixed value of $|a_0|$, 
and vanishes when noise is considerably 
high, as can be clearly seen from the figures.  
The regions (in Fig. \ref{gghz})  
marked by ``\textbf{R}'', and enclosed by the boxes are defined by 
the parameter ranges $0.65\leq|a_0|\leq0.7071$, and $0.4\leq p\leq0.6$, respectively. Note that in the marked areas, 
$\delta_\mathcal{N}>0$ for global noise, while $\delta_\mathcal{N}=0$ under DP channel. On the other hand, in the region \textbf{R}, 
$\delta_\mathcal{D}>0$ for both AD and PD channels. The implications of these values are discussed in Sec. \ref{sec:discrimination-tool}.

\subsection{Generalized W states}
\label{sec:gw}
Let us now move to the monogamy scores of negativity and quantum discord for the gW state, given by
\begin{equation}
|\Phi_3\rangle=a_0|001\rangle+a_1|010\rangle+a_2|100\rangle,
\label{eq:gw}
\end{equation}
where $a_0,a_1$ and $a_2$ are complex numbers, satisfying $|a_0|^2+|a_1|^2+|a_2|^2=1$. 
Unlike the gGHZ state, the reduced states obtained from the gW states with, as well as without, noise, 
are quantum correlated states, possessing non-zero values of entanglement as well as quantum discord.
This makes the evaluation of monogamy scores, for both negativity and quantum discord, involved, in comparison to the case of the gGHZ state.
However, analytical expressions for 
$\delta_\mathcal{N}$ can be determined for the global as well as the different types of local noise \cite{supp}. 
On the other hand, the computation of $\delta_{\mathcal{D}}$, in case of the gW states under noise,
requires more numerical resources than that in the case of the gGHZ states, since both 
$\mathcal{D}\big(\rho^{\mbox{\scriptsize gW\normalsize}}_{12}\big)$
and $\mathcal{D}\big(\rho^{\mbox{\scriptsize gW\normalsize}}_{13}\big)$ do not vanish for almost all  $p$. 
In the present case, $\delta_\mathcal{D}$ can be written as
\begin{eqnarray}
 \delta_\mathcal{D}=S-S(\rho^{\mbox{\scriptsize gW\normalsize}}_1)-S_c,
 \label{eq:deld_gw}
\end{eqnarray}
where $S=S(\rho^{\mbox{\scriptsize gW\normalsize}}_{12})+S(\rho^{\mbox{\scriptsize gW\normalsize}}_{13})
-S(\rho^{\mbox{\scriptsize gW\normalsize}})$, and $S_c=S(\rho^{\mbox{\scriptsize gW\normalsize}}_{2|1})
+S(\rho^{\mbox{\scriptsize gW\normalsize}}_{3|1})-S(\rho^{\mbox{\scriptsize gW\normalsize}}_{23|1})$.
The determination of $\delta_\mathcal{D}$ for a single three-qubit state requires, in principle, three separate optimizations
for the terms in $S_c$.  
However, information acquired via numerical analysis using constrained optimization 
\cite{num-err_group,chanda_freeze,disc_num_anal,const_opt} may result in considerable reduction of the computational complexity
(see \cite{supp} for a discussion).

\begin{figure}
 \includegraphics[scale=0.3]{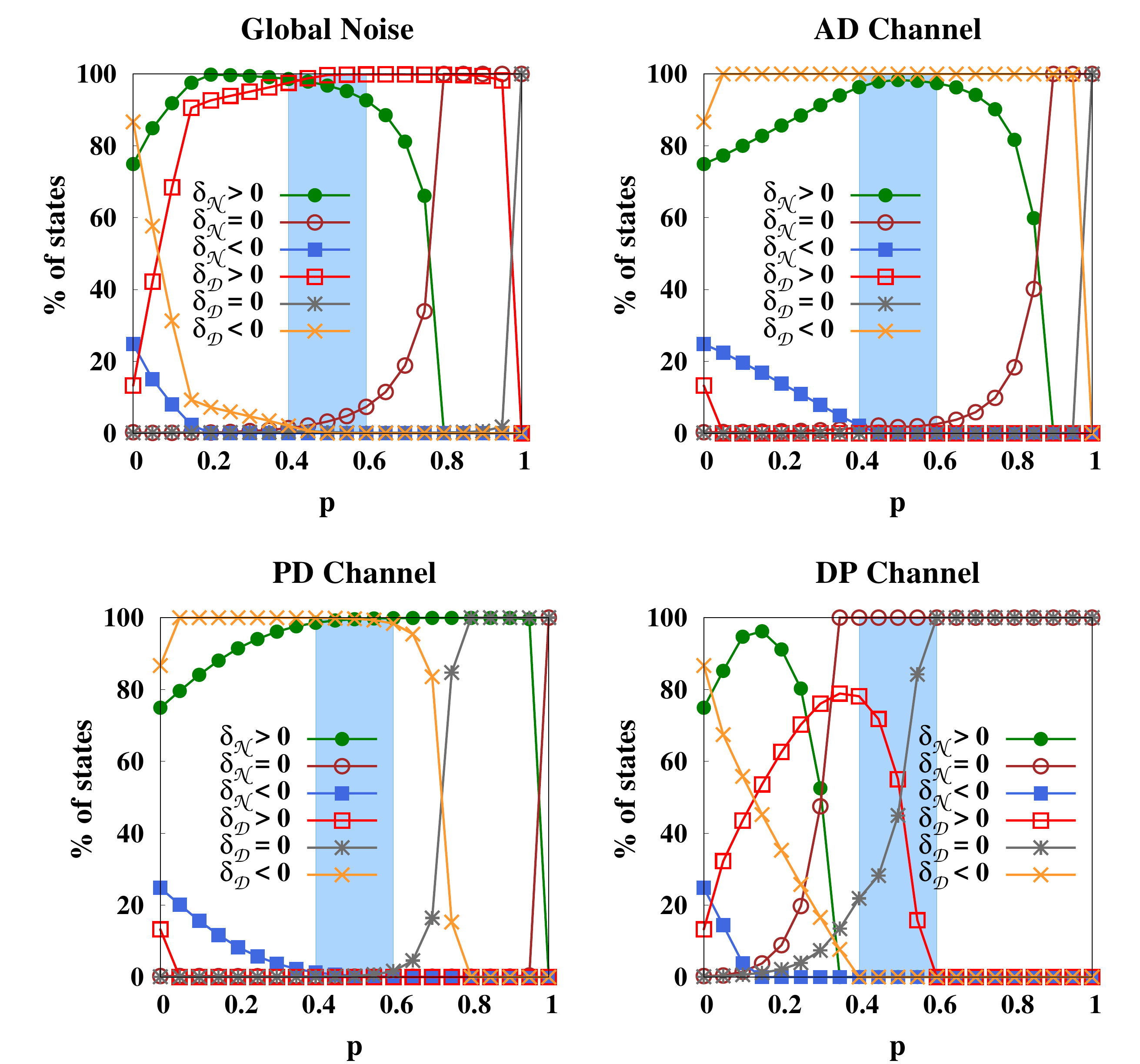}
 \caption{(Color online.) Variations of the percentages of states for which $\delta_{\mathcal{N}}$  and 
 $\delta_\mathcal{D}$ are greater than, equal to, and less than zero, for different types of noise 
 considered in this paper. The range of moderate noise, given by $0.4\leq p\leq0.6$, is shown by the shaded region in each figure. 
 All quantities plotted are dimensionless.}
 \label{gwper}
\end{figure}

\noindent{\textbf{Behavior of monogamy under moderate noise.}}
Let us now quantitatively study the behavior of monogamy scores, $\delta_\mathcal{N}$ and $\delta_\mathcal{D}$, 
of the gW states for a fixed noise parameter. 
We determine the fractions of the set of $\rho^{\mbox{\scriptsize gW\normalsize}}$, for which
the monogamy score corresponding to the chosen quantum correlation measure is strictly greater than, equal to, and strictly less than zero. 
We study the variation of these fractions with the change in values of the noise parameter for the specified type of noise. 
The variations of the three different fractions, as described above, with respect to $p$, are depicted in Fig. \ref{gwper}. 
Let us now investigate the effect of moderate noise on the monogamy scores. 
In the present study, we choose a 
range of $p$ given by $0.4\leq p\leq0.6$ (marked by the shaded regions in the panels in Fig. \ref{gwper}), 
which is moderate in comparison to the lower and upper bounds of $p$.   
From Fig. \ref{gwper}, it is clear that for moderate values of $p$ (viz. $0.4\leq p\leq0.6$), in the case of global noise as well 
as the AD channel, most of the states have $\delta_\mathcal{N}>0$, while for the DP channel, $\delta_\mathcal{N}=0$ 
for $100\%$ of the states. 
Remarkably, for the PD channel, all the states have $\delta_\mathcal{N}>0$ when the noise parameter is in the moderate range.

The situation is different for discord monogamy score. It is found that $\delta_\mathcal{D}\geq0$ for almost the 
entire range of moderate values of $p$, when gW states are subjected to global noise. In the case of the AD channel, 
$\delta_\mathcal{D}<0$ for the entire range $0.4\leq p\leq0.6$. In this scenario,  
$\delta_\mathcal{D}<0$ for the entire range of $p$, 
except only at $p=1$, the fully decohered states. Also, for the PD channel, $\delta_\mathcal{D}<0$ for moderate $p$ except 
when $p\approx0.6$. However, in the case of the DP channel, $\delta_\mathcal{D}\geq0$ for $0.4\leq p\leq0.6$. 
Hence it is clear that the monogamy of negativity behaves differently than the monogamy of quantum discord, in the case of 
global noise, and local channels considered in this paper. These results are of prime importance to our goal of 
channel discrimination, which will be discussed in Sec. \ref{sec:discrimination-tool}.

\noindent{\textbf{Types of dynamics.} 
In case of the gW state subjected to global or local noise, the set of different types of dynamics that $\delta_\mathcal{N}$ and 
$\delta_\mathcal{D}$ undergo is far richer compared to that for gGHZ states. While only monotonic decay of 
$\delta_{\mathcal{N}}$ and $\delta_\mathcal{D}$ with increasing $p$ is found in the latter case, non-monotonic dynamics of monogamy 
scores emerges in the former (see \cite{supp} for an example).  
Now we catalog four ``typical'' dynamics profiles observed for both $\delta_\mathcal{N}$ and $\delta_\mathcal{D}$ for global noise as 
well as for AD, PD, and DP local channels. 
\textbf{a.} In the first profile, $\delta_\mathcal{Q}(p=0)\geq0$, and $\delta_\mathcal{Q}(p)$ goes to 
zero non-monotonically as $p\rightarrow 1$. \textbf{b.} For the second one, 
$\delta_\mathcal{Q}(p)$ monotonically goes to zero when $p$ increases, with $\delta_\mathcal{Q}(p=0)\geq0$. 
\textbf{c.} In contrast to the first two profiles, $\delta_\mathcal{Q}(p=0)<0$ for the third profile. With an increase
of $p$, $\delta_\mathcal{Q}$ vanishes non-monotonically. \textbf{d.} Similar to the third profile, the 
fourth and the final profile starts with a non-monogamous scenario ($\delta_\mathcal{Q}(p=0)<0$). However, with increasing $p$, 
$\delta_\mathcal{Q}(p)$ goes to zero monotonically as $p\rightarrow 1$. Fig. \ref{robust} depicts two typical profiles of 
type-\textbf{(a)} and type-\textbf{(c)} dynamics for $\delta_\mathcal{N}$ and $\delta_\mathcal{D}$, respectively. A statistics on the 
occurrences of the four types of dynamics in the case of gW states is given in \cite{supp}.

\begin{figure}
 \includegraphics[scale=0.4]{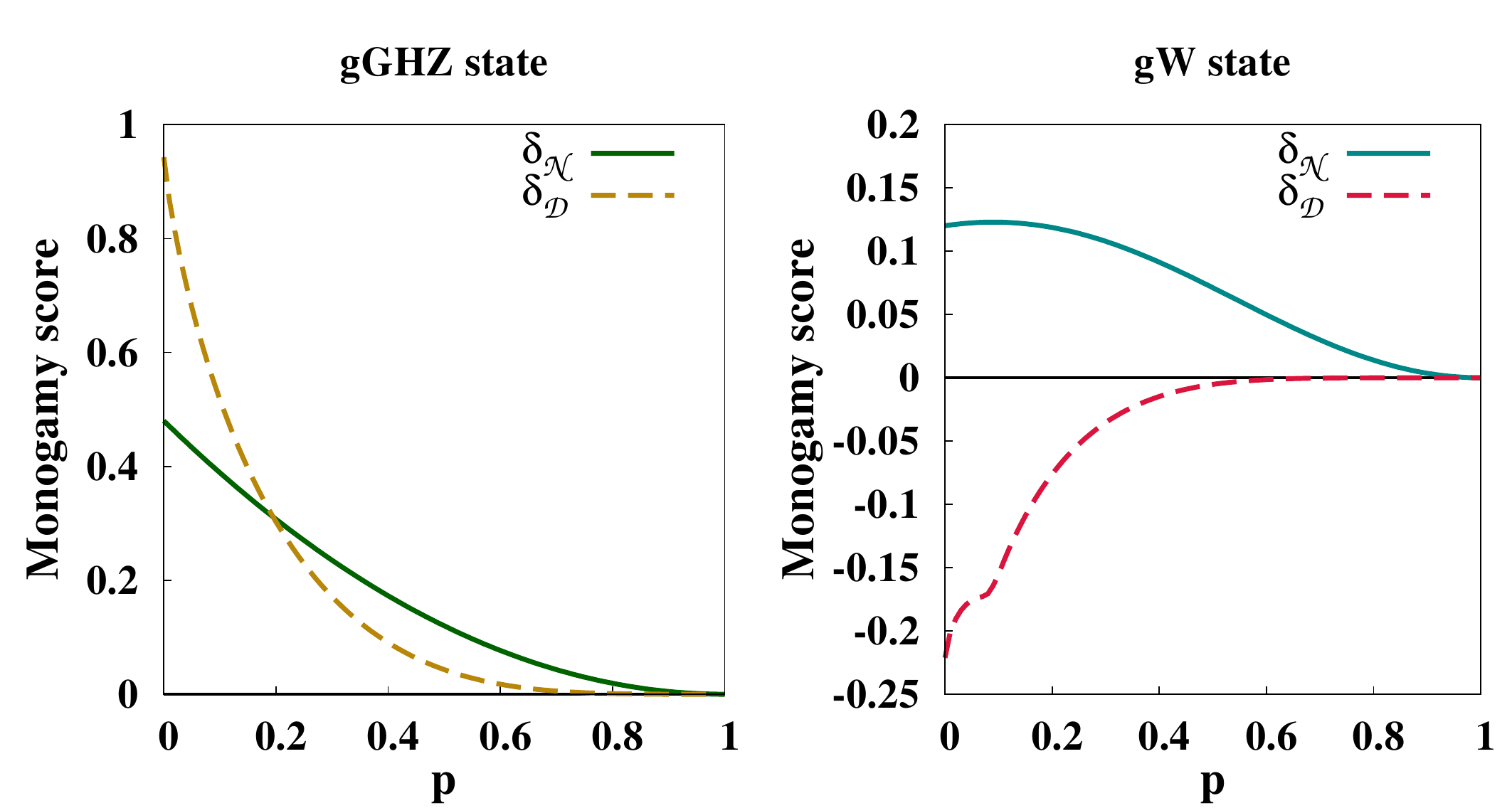}
 \caption{(Color online.) Dynamics profiles of $\delta_\mathcal{N}$ (solid line) and $\delta_\mathcal{D}$ (dashed line) 
 in the case of the input gGHZ state given by $|a_0|=0.7$ (left panel), and the input gW state given by $a_0=-0.287-0.552i$, 
 and $a_1=0.637+0.23i$ (right panel). In both the cases, $\delta_\mathcal{N}$ remains positive for higher values of $p$ ($p\geq0.6$),
 while $\delta_\mathcal{D}$ vanishes. Note that for the gW states considered here,  
 $\delta_\mathcal{N}$ exhibits a type-\textbf{a} dynamics, while that of $\delta_\mathcal{D}$ is of type-\textbf{c}.
 All the quantities plotted are dimensionless, except for $\delta_\mathcal{D}$, which is in bits.}
 \label{robust}
\end{figure}

\begin{table}
  \begin{tabular}{|c|c|c|}
  \hline
  Noise & $\langle p_t\rangle$ for $\delta_\mathcal{N}$ & $\langle p_t\rangle$ for $\delta_\mathcal{D}$ \\
  \hline 
  Global & $0.733$  & $0.947$  \\
  \hline
  AD & $0.667$  & $0.986$ \\
  \hline 
  PD & $0.940$  & $0.584$ \\
  \hline
  DP & $0.274$ & $0.331$ \\
  \hline  
  \end{tabular}
 \caption{The average values of dynamics terminal,  $\langle p_t\rangle$, for $\delta_\mathcal{N}$ and $\delta_\mathcal{D}$,
 when gW states are subjected to various kinds of noise. The profiles of the probability density function, $P(p_t)$,
 corresponding to different types of noise, for both $\delta_\mathcal{N}$ and $\delta_\mathcal{D}$, are given in 
 \cite{supp}.}
 \label{tab4}
 \end{table}

\noindent{\textbf{Dynamics terminal.}}  
So far, we have qualitatively discussed and characterized the dynamics of $\delta_\mathcal{N}$ and $\delta_\mathcal{D}$ 
under the application of global and local noise to the gW state. 
It is observed as well as intuitively clear that the persistence of the monogamy scores, when subjected to noise, must be 
different for different types of noise considered in this paper. To analyze this quantitatively, for a given state, $\rho$, we define 
the ``dynamics terminal'', $p_t$, which is given by the value of the noise parameter, 
$p$, at which the monogamy score vanishes, and remains so for $p_t\leq p\leq1$. The value of $p_t$ is 
characteristic to the input state, $|\Phi\rangle$, and the type of noise applied to it. A high value of $p_t$ implies 
a high persistence of the monogamy score for the state $|\Phi\rangle$ against the particular type of noise applied to it. 
It is clear that for gW states as the input states, $p_t$ may assume a range of values since the dynamics terminal will clearly 
have different values for different input gW states, when the type of noise is fixed.
However, for a specific type of noise, the average value of $p_t$, denoted by $\langle p_t\rangle$, and defined by
\begin{eqnarray}
 \langle p_t\rangle&=&\int_0^1p_t P(p_t)dp_t,
 \label{eq:pt_av}
\end{eqnarray}
provides a scale for the ``high'' values of the noise parameter. 
Here, $P(p_t)$ is the normalized probability density function (PDF) such that $P(p_t)dp_t$ provides the probability that for an arbitrary 
three-qubit gW state under the fixed type of noise, the value of $p_t$ lies between $p_t$ and 
$p_t+dp_t$. 
Note that the full range of the allowed values of $p_t$ is given by $0\leq p_t\leq1$, which follows from the 
definition of the noise parameter. 
The values of $\langle p_t\rangle$ corresponding to $\delta_\mathcal{N}$ and 
$\delta_\mathcal{D}$, calculated from Eq. (\ref{eq:pt_av}), for global noise, AD channel, PD channel, and DP channel are 
given in Table \ref{tab4} (see \cite{supp} for the numerical details). 

\noindent\textbf{Note.} Due to the extensive numerical effort required for 
determining the values of $\mathcal{D}(\rho^{\mbox{\scriptsize gW\normalsize}}_{1:23})$, 
$\mathcal{D}(\rho^{\mbox{\scriptsize gW\normalsize}}_{12})$, 
and $\mathcal{D}(\rho^{\mbox{\scriptsize gW\normalsize}}_{13})$ in computing $\delta_\mathcal{D}$ when $p>0$, 
we employ the constrained optimization technique \cite{supp,num-err_group,chanda_freeze,disc_num_anal,const_opt} 
to obtain several important statistics reported
in \cite{supp}. However, the error in the various statistics 
obtained for different channels, due to this approximation, is insignificant, and does not change the qualitative aspects of the results. 
Note that in all the occasions in this paper, where actual value of $\delta_\mathcal{D}$ has been plotted, or reported, exact optimization has 
been carried out using numerical techniques.

\noindent{\textbf{Robustness of negativity monogamy score.}}
As already mentioned in the Introduction, in the bipartite domain, it has been observed that quantum discord vanishes 
asymptotically with increasing noise strength, $p$, when quantum states are exposed to local noise. 
On the other hand, entanglement measures undergo a ``sudden death'' at a finite value of 
$p$ under similar noise, indicating a more fragile behavior than quantum discord. 
Interestingly, an opposite trend is observed when monogamy of quantum correlations are subjected to local noisy channels.
The variation of $\delta_\mathcal{N}$ and $\delta_\mathcal{D}$ with $|a_0|$ and $p$ in the case of PD channels with 
gGHZ states as input states (Fig. \ref{gghz}) indicates that there exists gGHZ states for which $\delta_\mathcal{N}$
persists longer than $\delta_\mathcal{D}$ for higher values of the noise parameter, $p$ ($p\geq0.8$). 
Our analysis also shows that the value of $\langle p_t\rangle$ for 
negativity monogamy score, in the case of the DP channel, is much larger compared to that of the discord monogamy score. 
Also, Fig. \ref{gwper} indicates that for higher values of $p$ ($0.6\leq p\leq0.9$), $100\%$ of gW states have 
$\delta_\mathcal{N}>0$ when the noise is of PD type. 
Note that for all such states, $\delta_\mathcal{D}\leq0$. This implies that there is a finite probability of finding gW states 
which, when subjected to PD channel, will evolve into a state $\rho^{\mbox{\scriptsize gW\normalsize}}$ with 
$\delta_\mathcal{N}>0$, but $\delta_\mathcal{D}=0$. 

We present two specific examples to establish such observations. Our first example is the gGHZ states represented by $|a_0|=0.6$, 
while the second example is the gW states given in Eq. (\ref{eq:gw}) with 
$a_0=-0.287-0.552i$, and $a_1=0.637+0.23i$. 
% where  $|a_2|$ is obtained from normalization. 
The behavior of $\delta_\mathcal{N}$ and $\delta_\mathcal{D}$ against $p$ are plotted in Fig. 
\ref{robust}, where the quantum discord components of $\delta_\mathcal{D}$ are computed via exact numerical optimization. 
It is clear from the figure that in both the cases, 
$\delta_\mathcal{N}$ persists longer than $\delta_\mathcal{D}$ at higher end of 
noise parameter. One must note here that the quantum discord components of 
$\delta_\mathcal{D}$ cancel each other at higher noise, while 
being individually non-zero. Hence the observation of $\delta_{\mathcal{N}}>0$ in situations where $\delta_{\mathcal{D}}=0$
is consistent with the fact that entanglement 
measures vanish for zero discordant states in bipartite systems. Therefore, it is evident that 
negativity monogamy score, in the presence of PD noise, exhibits a more robust behavior 
compared to that of the discord monogamy score. This is in contrast to the usual observation 
for bipartite quantum discord and entanglement measures.

Hitherto, we have investigated gGHZ and gW states, for which the effects of various noisy channels on monogamy scores can be addressed 
analytically up to certain extent. To complete the investigation for three-qubit states as input, we now consider the two
mutually exclusive and exhaustive classes of three-qubit  
states, viz., the GHZ class and the W class \cite{dvc}. 
These two classes, inequivalent under stochastic local operations and classical 
communication (SLOCC), together span the entire set of three-qubit pure states \cite{dvc}. We perform an analysis similar to the gW states, 
and find out how the percentages of states from each class, for which the monogamy score is greater than, equal to, or less than zero, 
varies with increasing noise parameter $p$ (see \cite{supp} for a detailed discussion). 
The patterns in the W class states are similar to those in the case of gW states, except for discord monogamy score under  
PD channel. While no gW states have a strictly positive $\delta_\mathcal{D}$ for higher values of $p$, in the case of W class states, the 
corresponding fraction increases with increasing $p$, reaches a maximum value at moderately high $p$, and then, as expected, 
decreases to zero as $p\rightarrow1$.

\begin{figure}
 \includegraphics[scale=0.35]{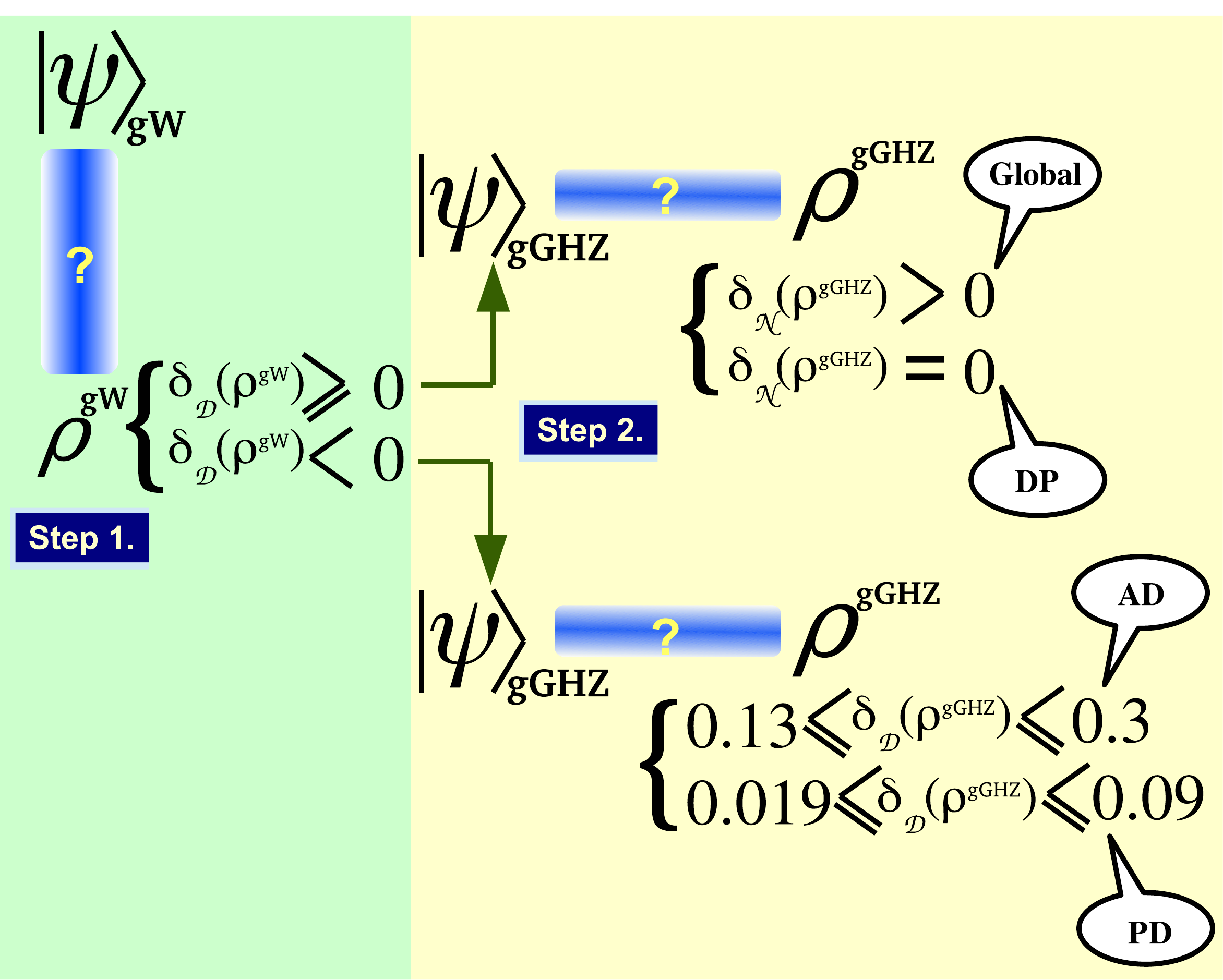}
 \caption{(Color online.) Schematic representation of the two-step channel discrimination protocol.}
 \label{protocol}
\end{figure}

\begin{figure}
 \includegraphics[scale=0.6]{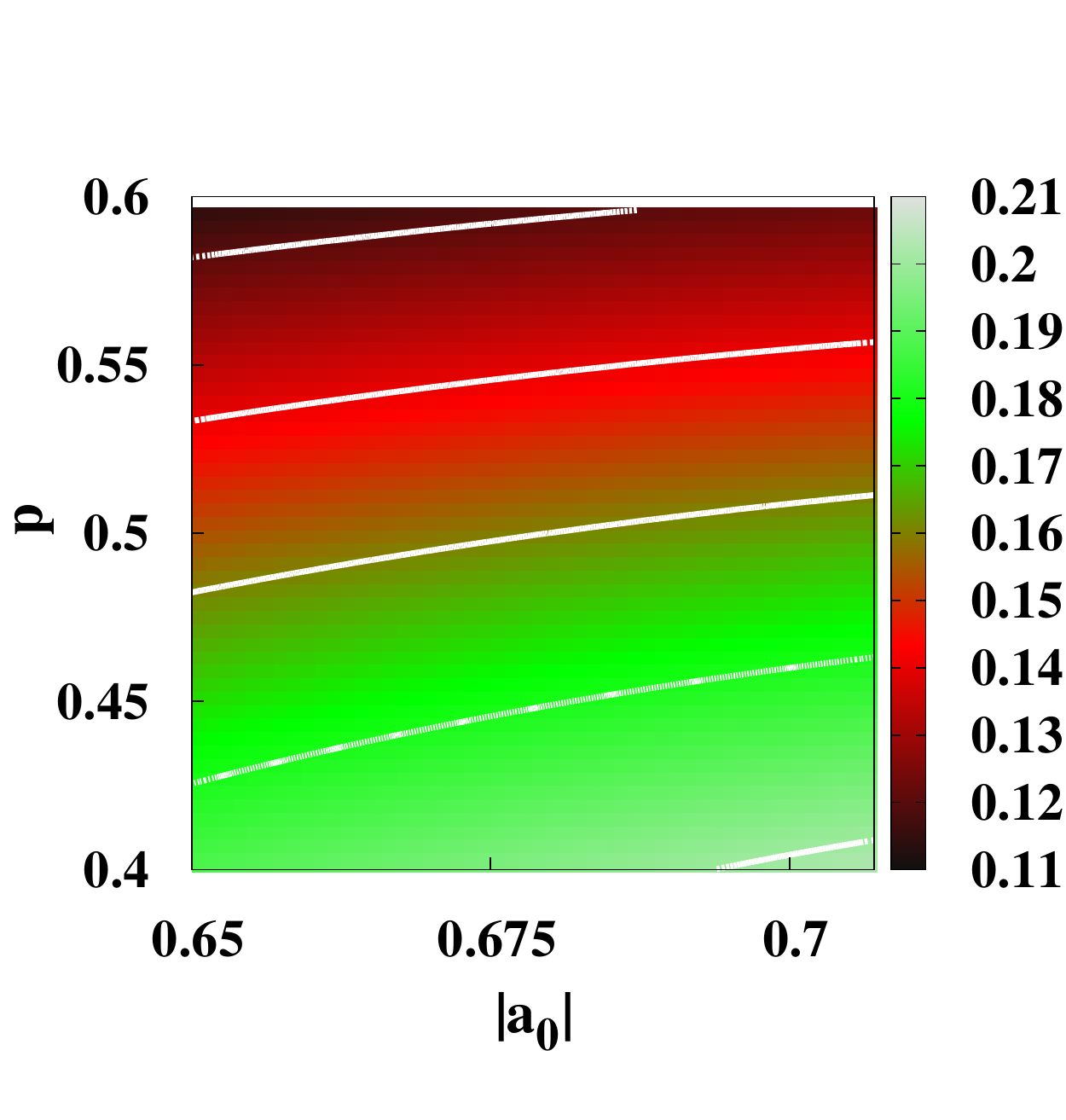}
 \caption{(Color online.) Variation of $\Delta$, the difference between the values of $\delta_\mathcal{D}$ for 
 a fixed gGHZ state under AD and PD channels, with the state parameter, $|a_0|$, and the noise parameter, $p$, in the 
 region ``\textbf{R}'' marked in Fig. \ref{gghz}. The solid lines are obtained by joining constant values of $\Delta$, where from 
 low to high value of $p$, the lines stand for $\Delta=0.12,0.14,0.16,0.18$, and $0.2$. All quantities plotted are dimensionless, except for 
 $\Delta$, which is in bits.}
 \label{gghz_enlarge}
\end{figure}

\section{Channel discrimination via monogamy}
\label{sec:discrimination-tool}
In this section, we investigate the second objective of this paper, and address the question whether 
monogamy of quantum correlations can be applied to conclusively detect the type of noise to which the quantum state is exposed. 
In particular, we propose a two-step protocol to discriminate global noise as well as local channels, namely, AD, PD, 
and DP channels, via negativity and discord monogamy score, by using a gW state and a gGHZ state as 
resources. The choice of observable in the second step is determined according to the outcome of the first step.  
The assumptions required for the success of the protocol are \textbf{(i)} that the strength of 
the noise is moderate, viz., $0.4\leq p\leq 0.6$, and \textbf{(ii)} that the given noisy channel can be used twice. 
Below, by an unknown channel, we shall mean one of th four channels, among global noise, AD, PD, and DP channels.

\noindent\textbf{Two-step discrimination protocol.} The two steps constituting the protocol are as follows. 
\textbf{1.} Given an unknown channel, the first step is to send an arbitrary gW state through that channel and to 
measure the value of $\delta_{\cal D}$ for the output state. \textbf{2.} The next step is to send a gGHZ state 
with high entanglement (e.g., $0.65\leq |a_0|\leq0.7071$) through the channel, and to measure 
the monogamy score corresponding to either negativity, or quantum discord, subject to 
the outcome of the first step. 
If $\delta_\mathcal{D}\geq0$ in the first step, $\delta_\mathcal{N}$ is chosen as the 
observable, while for $\delta_\mathcal{D}<0$ in step \textbf{1}, discord monogamy score can conclusively identify the 
type of noise in the channel (as shown schematically in Fig. \ref{protocol}).

\begin{table}[b]
 \begin{tabular}{|c|c|c|}
 \hline
  Step \textbf{1}: input gW  & Step \textbf{2}: input gGHZ & Conclusion \\ 
 \hline 
  $\delta_\mathcal{D}\geq0$ & $\delta_\mathcal{N}>0$ & Global noise\\
  \hline
  $\delta_\mathcal{D}<0$ & $0.13\leq\delta_{\mathcal{D}}\leq0.3$  & AD Channel\\
  \hline 
  $\delta_\mathcal{D}<0$ & $0.019\leq\delta_{\mathcal{D}}\leq0.09$ & PD Channel\\
  \hline 
  $\delta_\mathcal{D}\geq0$ & $\delta_\mathcal{N}=0$ & DP channel\\
  \hline 
 \end{tabular}
 \caption{Encoding of the outcomes of the two-step channel discrimination protocol using monogamy scores of negativity  
 and quantum discord.}
 \label{tab5}
\end{table}

Now we explain the implications of the output of the protocol. 
If $\delta_{\cal D}<0$ in step \textbf{1} for moderate values of \(p\), then the original gW state was subjected to either  
the AD, or the PD channel, while a nonnegative $\delta_{\cal D}$ implies that the noise was either global, or DP. This is 
clear from the variation of the percentages of states for which $\delta_\mathcal{D}\geq0$ and $<0$ in the range 
$0.4\leq p\leq0.6$, as depicted in Fig. \ref{gwper}. Hence, the first step divides the four types of noise in a block -- the duo of AD and PD 
channels, and global noise and DP channel. 

First, let us assume that $\delta_\mathcal{D}\geq0$ in the first step, which 
leads one to choose $\delta_{\cal N}$ as observable in the second step of the strategy.  
For $\delta_\mathcal{N}>0$ in the second step, the type of noise that acts on the gGHZ state is the global noise, 
while $\delta_{\cal N}=0$ implies that the channel is DP. This can be understood from the boxed regions marked 
``\textbf{R}'' in Fig. \ref{gghz}, where $\delta_\mathcal{N}=0$ for the DP channel,
while $\delta_\mathcal{N}>0$ for global noise. 
On the other hand, if the outcome of the first step is $\delta_\mathcal{D}<0$, the channel is either AD, or PD. 
In this situation, $\delta_\mathcal{D}$ is always positive when the input state is the gGHZ state with a specific value of $|a_0|$ in the 
range mentioned before, and so the discrimination protocol is more involved. In particular,   
we observe that in the marked region ``\textbf{R}'' in Fig. \ref{gghz}, $0.13\leq\delta_{\mathcal{D}}^{ad}\leq0.3$ for the 
AD channel, while for the PD noise, $0.019\leq\delta_\mathcal{D}^{pd}\leq0.09$. The variation of 
$\Delta=\delta_{\mathcal{D}}^{ad}-\delta_{\mathcal{D}}^{pd}$, the difference between the values of discord monogamy 
score in the case of AD channel ($\delta_{\mathcal{D}}^{ad}$) and PD channel ($\delta_{\mathcal{D}}^{pd}$),
with $|a_0|$ and $p$ in the region ``\textbf{R}'' in Fig. \ref{gghz} is plotted in Fig. \ref{gghz_enlarge}.
We notice that there is no overlap between the allowed ranges of $\delta_\mathcal{D}$ for the two channels (as also indicated by 
the absence of the value $\Delta=0$ in Fig. \ref{gghz_enlarge}), implying that $\delta_\mathcal{D}$ can 
conclusively distinguish between the AD and the PD channels. The possible encoding of the outcomes of the two-step protocol, and 
their implications are tabulated in Table \ref{tab5}.

\noindent\textbf{Remark 1.} The first step of our channel discrimination protocol requires not  
the value, but only the sign of $\delta_{\cal D}$, while the second step requires an estimation of the discord monogamy score, 
$\delta_\mathcal{D}$, for AD and PD channels.

\noindent\textbf{Remark 2.} Although the range of values of $\delta_\mathcal{D}$ are non-overlapping for AD and PD 
channels when the state parameter is in the range $0.6<|a_0|<0.7071$, the difference between the values corresponding to
the lower bound of $\delta_{\mathcal{D}}^{ad}$, and upper bound of $\delta_\mathcal{D}^{pd}$ decreases with relaxing the lower
bound of $|a_0|$. Hence, the lower bound of the allowed range of $|a_0|$ can be relaxed depending on the accuracy with which 
$\delta_\mathcal{D}^{ad}$ and $\delta_\mathcal{D}^{pd}$ can be estimated with the current technology in hand. The best 
result is obtained for the three-qubit GHZ state, for which $|a_0|=1/\sqrt{2}$.

\noindent\textbf{Remark 3.} In the presence of high noise ($p>0.6$), our protocol may fail to distinguish the type of 
noise applied to the quantum state. This is because both $\delta_{\cal D}$ and $\delta_\mathcal{N}$ may vanish in the 
case of both gGHZ state and gW state when the noise strength is high. It is also clear that the above distinguishing 
protocol fails when \(p\approx 0\).

\section{conclusion}
\label{sec:summary}

The question that naturally arises from the results reported in the preceeding sections is 
whether the dynamical features found in the case of three-qubit systems is generic for quantum states with higher number of qubits. The study
of the dynamical properties of monogamy scores in the multipartite domain with more than three parties is challenging due to computational
as well as analytical
difficulties in computing the correlation measures. In particular, the number of inequivalent SLOCC classes with higher number 
of parties becomes much higher \cite{dvc}. 
As we have already mentioned, the 
monogamy score of gGHZ state decreases monotonically with  the increase of $p$ for an arbitrary number of parties.  
And, the non-monotonic behaviour of the monogamy scores corresponding to both negativity and quantum discord,
in the case of the three-qubit generalized W state under, for example, the AD channel, is found in the case of higher number of qubits
(see \cite{supp}). 
We also find that similar to the three-qubit scenario, the depolarizing channel destroys the monogamy scores corresponding to both 
negativity and quantum discord more rapidly compared to other channels, for four qubits. We therefore expect that
the broad qualitative features  of the results
found in the case of three qubits remain unaltered when the number of qubits is increased.

Summarizing, we have investigated the patterns of the monogamy property of quantum correlations using monogamy 
score as the observable, when three-qubit 
systems are subjected to global noise as well as local noisy channels, viz. amplitude-damping, phase-damping, and depolarizing channels.
As the quantum correlation measures that are used to constitute the monogamy score, 
we chose negativity and quantum discord, and found that the dynamics of monogamy score, when generalized GHZ states 
are subjected to different types of noise, is qualitatively different from that of the generalized W state as input. While monogamy
score corresponding to both the quantum correlation measures exhibit a monotonic decay with increasing noise in the former case, non-monotonic 
dynamics takes place in the latter, giving rise to a rich set of dynamical profiles. 
We define a characteristic noise scale, called the ``dynamics terminal'', that quantifies the 
persistence of the monogamy score corresponding to a particular measure of quantum correlation, when the state is subjected to a specific 
type of noise. We show that the dynamics terminal can distinguish between the different noise models, and indicates that the depolarizing channel 
destroys monogamy scores faster compared to the other types of noise. To investigate how the monogamy property behaves against increasing noise, we 
investigate the variation of the fraction of states 
with increasing value of the noise parameter, when the input states are chosen from the GHZ and the  W class. 
We also show that the negativity monogamy score may exhibit a more robust behavior against phase damping noise, compared 
to the discord monogamy score, which is in contrast to the usual observation regarding bipartite entanglement measures 
and quantum discord. 
As an usefulness of such study, we propose a 
two-step channel discrimination protocol that can conclusively identify the different types of noise by considering 
monogamy scores and by using the gGHZ and the gW states as resources.

\begin{acknowledgments}
RP acknowledges support through an INSPIRE-faculty position at the
Harish-Chandra Research Institute, by the Department of Science and
Technology, Government of India. 
\end{acknowledgments}

% \pagebreak

\widetext
\begin{center}
\textbf{\Large Supplementary Materials\\ \large Conclusive Identification of Quantum Channels via Monogamy of Quantum Correlations} \\
\normalsize Asutosh Kumar\(^{1,2}\), Sudipto Singha Roy\(^{1,2}\), Amit Kumar Pal\(^{1,2}\), R. Prabhu\(^{1,2,3}\), Aditi Sen(De)\(^{1,2}\), and Ujjwal Sen\(^{1,2}\)\\
\small \(^1\)Harish-Chandra Research Institute, Chhatnag Road, Jhunsi, Allahabad 211019, India \\
\(^2\)Homi Bhabha National Institute, Training School Complex, Anushaktinagar, Mumbai 400094, India\\
\(^3\)Department of Physics, Indian Institute of Technology Patna, Bihta 801103, Bihar, India​
\end{center}

\setcounter{equation}{0}
\setcounter{figure}{0}
\setcounter{table}{0}
\setcounter{section}{0}
\setcounter{page}{1}
\makeatletter
\renewcommand{\theequation}{SEQ\arabic{equation}}
\renewcommand{\thefigure}{SF\arabic{figure}}
\renewcommand{\thetable}{ST\arabic{table}}
\renewcommand{\bibnumfmt}[1]{[SR#1]}
\renewcommand{\citenumfont}[1]{SR#1}
\renewcommand{\thesection}{SSEC\arabic{section}}
\hypersetup{pageanchor=false}

\section{Measures of quantum correlations}
\label{sec:qcmeasures}
In this paper, we restrict ourselves to 
two specific measures, namely, negativity and quantum discord. The first one belongs to the 
quantum correlations defined in the entanglement-separability domain, 
while the second one is an information-theoretic quantum correlation measure.

\noindent\textbf{\textit{Negativity.}}  
For a bipartite state \(\rho_{AB}\), its negativity \cite{S_neg_gr}, \({\cal N}(\rho_{AB})\), is defined as the absolute value of 
the sum of the negative eigenvalues of \(\rho_{AB}^{T_{A}}\), where \(\rho_{AB}^{T_{A}}\) denotes the partial transpose \cite{S_neg_part}
of \(\rho_{AB}\) with respect to the subsystem \(A\).
Alternatively, it is expressed as
 \begin{equation}
  {\cal N}(\rho_{AB})=\frac{\|\rho_{AB}^{T_A}\|_1-1}{2},
  \label{eq:negativity}
 \end{equation}
where $\|M\|_1 \equiv \mbox{tr}\sqrt{M^\dag M}$ is the trace-norm of the matrix $M$.

\noindent\textbf{\textit{Quantum Discord.}}
Quantum discord \cite{S_disc_group} is defined as the difference between the ``total correlation'' \cite{S_total_corr} 
and the ``classical correlation'' \cite{S_disc_group} present in the composite system, described by the bipartite state \(\rho_{AB}\).  
The total correlation can be quantified as the quantum mutual information, and is given by 
\begin{equation}
\mathcal{I}(\rho_{AB})= S(\rho_A)+ S(\rho_B)- S(\rho_{AB}),
\label{mutual_info}
\end{equation}
where $S(\varrho)= - \mbox{tr} (\varrho \log_2 \varrho)$ is the von Neumann entropy of $\varrho$, and $\rho_{A(B)}$
are the local density matrices of $\rho_{AB}$, obtained as $\rho_{A(B)}=\mbox{tr}_{B(A)}\left[\rho_{AB}\right]$. 
On the other hand, the classical correlation is defined as 
\begin{equation}
 {\cal J}(\rho_{AB}) = S(\rho_B) - S(\rho_{B|A}),
 \label{class_corr}
\end{equation}
where the conditional entropy, $S(\rho_{B|A})$, is given by
\begin{equation}
 S(\rho_{B|A}) = \min_{\{P_i\}} \sum_i p_i S(\rho_{B|i}).
\end{equation} 
Here, $S(\rho_{B|A})$ is conditioned over measurement performed on \(A\) with a rank-one projection-valued measurements \(\{P_i\}\),
producing the states  
\(\rho_{B|i} = \frac{1}{p_i} \mbox{tr}_A[(P_i \otimes \mathbb{I}_B) \rho_{AB} (P_i \otimes \mathbb{I}_B)]\), 
with probability \(p_i = \mbox{tr}[(P_i \otimes \mathbb{I}_B) \rho_{AB} (P_i \otimes \mathbb{I}_B)]\),  
\(\mathbb{I}_B\) being the identity operator in the Hilbert space of \(B\). From Eqs. (\ref{mutual_info}) and 
(\ref{class_corr}), quantum discord can be obtained as  
\begin{equation}
{\cal D}(\rho_{AB})= {\cal I}(\rho_{AB}) - {\cal J}(\rho_{AB}).
\label{qdiscord}
\end{equation} 

\noindent\textbf{Note.} The difficulty in the computation of quantum discord arises due to 
the optimization involved in the definition of classical 
correlation of the state $\rho_{AB}$ \cite{S_huang_disc_np}. In the case of a pure bipartite state $\rho_{AB}$, quantum discord 
reduces to $S(\rho_{A})$, the von Neumann entropy of the local density matrix $\rho_A$ \cite{S_koashi}. On the other hand, there are only a few 
examples of mixed bipartite states, for which quantum discord can be obtained analytically \cite{S_luo_pra_2008,S_disc_2q}. 
For an arbitrary mixed bipartite state 
$\rho_{AB}$, computation of quantum discord involves adaptation of numerical 
optimization techniques \cite{S_num-err_group}. 
In the case of a 
$\mathbb{C}^2\otimes\mathbb{C}^d$ system, if measurement is performed on the qubit, the
rank-\(1\) projectors, $\{P_i=|\Phi_i\rangle\langle\Phi_i|,i=1,2\}$, can be parametrized as
\begin{eqnarray}
|\Phi_1\rangle&=&\cos \frac{\theta}{2}|0\rangle + e^{i\phi}\sin \frac{\theta}{2}|1\rangle, \nonumber \\
|\Phi_2\rangle&=&-e^{-i\phi}\sin\frac{\theta}{2}|0\rangle + \cos \frac{\theta}{2}|1\rangle.
\end{eqnarray} 
The optimization, in this case, is to be performed over the space of the real parameters $(\theta,\phi)$, where 
$0 \leq \theta \leq \pi$ and $0 \leq \phi < 2\pi$.

\section{Monogamy of Quantum Correlations}
\label{sec:monogamy}

Let us now introduce the concept of monogamy \cite{S_mono} for a quantum correlation measure, $\mathcal{Q}$. 
Monogamy helps in exploring the amount of quantum correlations shared among the subsystems of a quantum system. 
All the known quantum correlation measures qualitatively follow a monogamy relation. 
In the case of a tripartite system, it implies that if two of the parties are maximally quantum correlated, then there can not 
be any quantum correlation between either of these two parties with the third one. 
Importantly, classical correlations do not have such restrictions. 
We will now quantify the monogamy constraint for an arbitrary bipartite quantum correlation measure, 
say ${\cal Q}$. An \(n\)-party state, $\rho_{A_1A_2\cdots A_n}$, shared between the parties, $A_1,A_2,\cdots,A_n$, is said to 
be monogamous under the quantum correlation measure ${\cal Q}$, if it follows the monogamy inequality given by
\begin{equation}
 {\cal Q}(\rho_{A_1:A_2\cdots A_n})\geq \sum_{j=2}^n{\cal Q}(\rho_{A_1A_j}),
\end{equation}
where $\rho_{A_1A_j}$ is obtained from $\rho_{A_1A_2\cdots A_n}$ by tracing out all the parties except $A_1$ and $A_j$.  
Otherwise, it is non-monogamous. Here we call the party $A_1$ as the nodal observer. In this respect, the ``monogamy 
score'' with respect to \({\cal Q}\),
for the \(n\)-party state, $\rho_{A_1A_2\cdots A_n}$, is defined as \cite{S_mono_score}
\begin{equation}
\delta_{{\cal Q}} = {\cal Q}(\rho_{A_1:A_2\cdots A_n}) - \sum _{j=2}^{n}{\cal Q}(\rho_{A_1A_j}). 
\label{eq:monoscore}
\end{equation}
Therefore, positivity of \(\delta_{{\cal Q}}\) for a given quantum state implies 
monogamy of quantum correlation measure \({\cal Q}\) for that state.

\noindent\textbf{Note.} The choice of measurement, in the definition of quantum discord, puts an inherent asymmetry in the measure. 
In this paper, unless otherwise stated, the measurement is performed on the first subsystem of the bipartite quantum system. 
This implies that while computing $\mathcal{D}(\rho_{A_1A_j})$ to determine $\delta_{\mathcal{Q}}$, 
the measurement is always performed on the nodal observer.

\section{Decoherence under global and local noise}
\label{sec:noise-models}

A quantum system inevitably interacts with its environment and eventually decoheres, and looses its quantum correlations.
Such a decohering process can be described by a completely positive trace preserving (CPTP) map, $\mathcal{E}$, which,
acting on the quantum system, $\rho$, transforms the state as 
\cite{S_decohere_rev,S_mazeiro_disc_dyn,S_operator-sum,S_kraus,S_preskill,S_sudarshan,S_choi}
\begin{equation}
\rho \rightarrow \rho^\prime={\cal E}(\rho).
\end{equation}
The noise can act either globally, or locally on 
each subsystem of the system of interest. In the present study, we consider both the scenarios.

\noindent\textbf{\textit{Global noise.}}
In this case, we consider an environment that acts globally on a system of dimension $d^n$ and in the state $\rho$, for which the
resulting state is given by
\begin{equation}
\rho^\prime=\frac{p}{d^n}I+(1-p)\rho,
\label{eq:global}
\end{equation}
where \(p\) is the mixing parameter $(0\leq p\leq 1)$, and $I$ is the identity operator in the Hilbert space of the system.
Note that $p=0$ stands for the noiseless case, while $p=1$ corresponds to the fully decohered state. 

\noindent\textbf{\textit{Local noise.}} 
For a composite quantum system having $n$ spatially separated subsystems, it is reasonable to assume 
that the environment acts independently and locally on each of the subsystems. 
We now briefly describe various such local noisy channels. 

The dynamics of a closed quantum system is described by a 
unitary transformation. To describe the dynamics of an open quantum system, which is interacting with its environment, 
one can assume that the system and the environment together form a closed quantum system, whose state, $\rho$, 
is given by $\rho=\rho_s\otimes\rho_e$, with  $\rho_s$ and 
$\rho_e$ being respectively the states of the system and the environment. The next step would be to apply a unitary transformation 
to the given composite system and finally trace out the environment part to obtain the reduced state of the changed system.
In this case, quantum operations can be considered in the \textit{operator-sum representation} \cite{S_operator-sum}, 
written explicitly in terms of operators on the Hilbert space of the system as follows:
\begin{equation}
\rho^\prime_s=\mbox{tr}_{e} [U(\rho_s\otimes\rho_{e})U^{\dagger}]=\sum_k E_k\rho_sE_k^{\dagger},
\end{equation}
where the operators {$\{E_k\}$} are known as Kraus operators \cite{S_mazeiro_disc_dyn,S_operator-sum,S_kraus,S_preskill} 
and satisfy $\sum_k E_k^{\dagger}E_k = I$.
For a system ``$s$'' of dimension $d$, any quantum operation can be represented by at most $d^2$ Kraus operators. 

For an \(n\)-partite system, \(\rho_{A_1A_2\cdots A_n}\), in arbitrary dimensions, after the actions of the local environments 
on the subsystems, the evolved state, 
$\rho^\prime_{A_1A_2\cdots A_n}$, can be written as 
\begin{equation}
\rho^\prime_{A_1A_2\cdots A_n}=\sum_{k_1,k_2,\cdots,k_n}E_{k_1k_2\cdots k_n}\rho_{A_1A_2\cdots A_n}E_{k_1k_2\cdots k_n}^\dagger,
\label{eq:multiparty_evolved}
\end{equation} 
with $E_{k_1k_2\cdots k_n}=E_{k_1}^{(1)}\otimes E_{k_2}^{(2)}\otimes \cdots \otimes E_{k_n}^{(n)}$.  
Here, $E_{k_j}^{(j)}$, $j=1,2,\cdots,n$, are the Kraus operator for the local action on subsystem $A_{j}$ with dimension 
$d_j$ so that $0\leq k_j\leq d_j^2-1$. 
Now, we describe the Kraus operators of a number of single-qubit quantum channels, namely, 
the amplitude- and phase-damping, and depolarizing channels \cite{S_mazeiro_disc_dyn,S_preskill}. 

\noindent\textit{Amplitude-damping channel.}  
The AD channel represents a scenario where energy dissipation 
from a quantum system is allowed. 
The Kraus operators for a single-qubit AD channel are given by
\begin{eqnarray}
  E_0=\left(
  \begin{array}{cc}
     1 & 0\\
     0 & \sqrt{1-p}
  \end{array}
  \right),~~~
  E_1=\left(
  \begin{array}{cc}
     0 & \sqrt{p}\\
     0 & 0
  \end{array}
  \right),
  \label{eq:ad_kraus}
\end{eqnarray}
with $0\leq p\leq 1$, where $p$ corresponds to the strength of the noise acting on the input qubit state.

\noindent\textit{Phase-damping channel.}
As an example of a non-dissipative channel, we consider the PD channel. 
A state, after passing through the PD channel, or the ``dephasing'' channel, decays its 
off-diagonal elements, resulting in information loss about its coherence. 
The single qubit Kraus operators for the PD channel are given by
\begin{equation}
E_0=\sqrt{1-p}\textit{I},~E_1=\frac{\sqrt{p}}{2}(\textit{I}+\sigma_3), ~E_2=\frac{\sqrt{p}}{2},(\textit{I}-\sigma_3),
\label{eq:pd_kraus}
\end{equation}
where $I$ is the identity matrix in the qubit Hilbert space, and $p$ is again the noise-strength. 

\noindent\textit{Depolarizing channel.}
In the case of the  DP channel, the input qubit is depolarized, that is, replaced by the completely mixed state $\frac{\textit{I}}{2}$, 
with probability \(p\) and is left unaltered with probability \((1-p)\). Such an operation on the single-qubit state $\rho$ is 
represented by 
\begin{equation}
\label{eq:dpc}
\rho^\prime=\frac{p}{2}\textit{I}+(1-p)\rho.
\end{equation}
Note that the form in Eq. (\ref{eq:dpc}) is not in the operator-sum representation. The operation given in Eq. (\ref{eq:dpc}) 
is often parametrized as
\begin{equation}
\rho^\prime=(1-p)\rho+\frac{p}{3}(\sigma_1\rho\sigma_1+\sigma_2\rho\sigma_2+\sigma_3\rho\sigma_3),
\end{equation}
leading to single qubit Kraus operators of the form given by 
\begin{equation}
E_0=\sqrt{1-p}\textit{I},~E_1=\sqrt{\frac{p}{3}}\sigma_1,~E_2=\sqrt{\frac{p}{3}}\sigma_2,~E_3=\sqrt{\frac{p}{3}}\sigma_3.
\label{eq:dp_kraus}
\end{equation}

\noindent\textbf{Remark.} Similar to the case of global noise, in the case of local noisy channels also, the noiseless case is 
denoted by $p=0$, while $p=1$ represents maximal disturbance of the state.

\section{Negativity monogamy score under decoherence}
\label{sec:neg-score}

\subsection{Generalized GHZ state}
\label{subsec:gghz-neg-score}

Let us first consider the case of global noise acting on the gGHZ state. The final state, 
$\rho^{\mbox{\scriptsize gGHZ\normalsize}}$, as a function of the mixing parameter, $p$, can be obtained following the methodology 
described in \ref{sec:noise-models}, 
which leads to two-party reduced states $\rho_{12}^{\mbox{\scriptsize gGHZ\normalsize}}$ and 
$\rho_{13}^{\mbox{\scriptsize gGHZ\normalsize}}$ of the form
$\rho^{\mbox{\scriptsize gGHZ\normalsize}}_{12}=\rho^{\mbox{\scriptsize gGHZ\normalsize}}_{13}
=(1-p)(|a_0|^2|00\rangle \langle 00|+|a_1|^2|11\rangle \langle 11|)+\frac{p}{4}I$,
with $I$ being a $4\times 4$ identity matrix. They still remain 
classically correlated with vanishing entanglement and quantum discord. 
In case of AD, PD, and DP channels, the resulting states $\rho^{\mbox{\scriptsize gGHZ\normalsize}}$ are obtained as 
(see \ref{sec:noise-models})
$\rho^{\mbox{\scriptsize gGHZ\normalsize}}=\sum_{i=0}^{1}|a_i|^2\left(u^p_i|0\rangle\langle 0|+v^p_i|1\rangle\langle 1|\right)^{\otimes 3}
 +w^p(a_0a_1^*|000\rangle\langle111|+h.c.)$.
Here the functions $u^p_i$, $v^p_i$, and $w^p$, for the three channels, are given by
\begin{eqnarray}
 \mbox{\small AD channel\normalsize} &:& u^p_i=\delta_{0i}+p\delta_{1i},v^p_i=(1-p)\delta_{1i},
 w^p=(1-p)^{\frac{3}{2}},\nonumber \\
 \mbox{\small PD channel\normalsize} &:& u^p_i=\delta_{0i},v^p_i=\delta_{1i},w^p=(1-p)^3,\nonumber \\
 \mbox{\small DP channel\normalsize} &:& u^p_i=q\delta_{0i}+(1-q)\delta_{1i},
                                         v^p_i=(1-q)\delta_{0i}+q\delta_{1i},\nonumber \\
                                         && w^p=(2q-1)^3,
\end{eqnarray}
with $q=1-\frac{2p}{3}$. From the above expressions, it can be shown that 
the two-qubit reduced density matrices, in case of the PD
channel, do not depend on the noise parameter $p$, and remain classically correlated. On the other hand, 
$\rho^{\mbox{\scriptsize gGHZ\normalsize}}_{12}$ and $\rho^{\mbox{\scriptsize gGHZ\normalsize}}_{13}$ remains diagonal in the computational basis 
$\{|00\rangle,|01\rangle,|10\rangle,|11\rangle\}$, resulting in vanishing entanglement as well as quantum discord
for the other two channels also. 
In effect, even for $p\neq 0$, $\delta_{\mathcal{N}}$ and $\delta_{D}$ are given by 
$\delta_{\mathcal{N}}=\mathcal{N}\big(\rho^{\mbox{\scriptsize gGHZ\normalsize}}_{1:23}\big)$ and 
$\delta_\mathcal{D}=\mathcal{D}\big(\rho^{\mbox{\scriptsize gGHZ\normalsize}}_{1:23}\big)$ respectively,
when the gGHZ state is subjected to these four types of noise.
Hence, both negativity and quantum discord are always monogamous in the present scenario, which can be applied to 
discriminate channels. 
% as we shall see in Sec. \ref{sec:discrimination-tool}. 
% Note here that $S(\rho^\prime_{23|1})$ is computed by performing measurement over the nodal observer (qubit $1$). 
Note that all the above discussions hold for the gGHZ state of arbitrary number of parties
subjected to different types of local and global noise considered in this paper.

Using the expression for $\rho^{\mbox{\scriptsize gGHZ\normalsize}}$, analytical expressions of  $\delta_{\mathcal{N}}$, as functions
of the noise parameter, $p$, and the state parameter, $|a_0|$, can be obtained for different types of noise. 
In the case of the global noise, it is given by 
\begin{eqnarray}
 \delta_{{\cal N}}^{g} &=& \Big|\min\Big[0,\frac{1}{2}\Big\{\frac{p}{4}-2|a_0||a_1|(1-p)\Big\}\Big]\Big|,
\end{eqnarray}
while in the case of PD channel,  
\begin{eqnarray}
 \delta_{\mathcal{N}}^{pd}&=&|a_0||a_1|(1-p)^3.
\end{eqnarray}
The expressions of negativity monogamy score in the case of AD channel ($\delta_{\mathcal{N}}^{ad}$) and DP 
channel ($\delta_\mathcal{N}^{dp}$) are given by  
\begin{eqnarray}
\delta_{\mathcal{N}}^{ad}&=&\Big|\min
\Big[0,\frac{1}{2}\Big\{|a_1|^2p(1-p)-\sqrt{f_{1}^{ad}+f_2^{ad}}\Big\}\Big]\Big|,\\
\delta_{\mathcal{N}}^{dp}&=&\Big|\min\Big[0,\frac{1}{2}\Big\{q(1-q)-\sqrt{f_{1}^{dp}+f_{2}^{dp}}\Big\}\Big]\Big|,
\label{eq:neg_score}
\end{eqnarray}
with the functions $f_1^{ad}$, $f_{2}^{ad}$, $f_{1}^{dp}$, and $f_{2}^{dp}$ defined as 
$f_1^{ad}=|a_1|^4(4p^6-12p^5+13p^4-6p^3+p^2)$,
$f_2^{ad}=4|a_0|^2|a_1|^2(1-p)^3$,
$f_1^{dp}=q^2(1-q)^2(1-2q)^2$, and
$f_2^{dp}=4|a_0|^1|a_1|^2(1-2q)^2(1-3q+3q^2)(1-5q+5q^2)$.
Note here that in all the above expressions, one can replace $|a_1|$ by $\sqrt{1-|a_0|^2}$.

\subsection{Generalized W state}
\label{gw-neg-score}

When the gW state is subjected to global noise, 
the evolved three-qubit state, $\rho^{\mbox{\scriptsize gW\normalsize}}$
leads to the two-qubit reduced density matrix, $\rho_{12}^{\mbox{\scriptsize gW\normalsize}}$, of the form 
$\rho^{\mbox{\scriptsize gW\normalsize}}_{12}=\frac{p}{4}I_4+(1-p)\left(|a_0|^2P[|00\rangle]+P[|\psi\rangle]\right)$
in the computational basis, where $P[|x\rangle]=|x\rangle\langle x|$, and $|\psi\rangle=a_1|01\rangle+a_2|10\rangle$. 
The reduced state of qubits $1$ and $3$ can be determined from 
$\rho^{\mbox{\scriptsize gW\normalsize}}_{12}$ by interchanging $a_0$ and $a_1$.  
One should note here that unlike the gGHZ state, the reduced states with, as well as without, noise in the current case, 
are no more ``classical-classical'' states, and possess non-vanishing entanglement as well as quantum discord.  
For the AD channel, the three-qubit resulting  state,
starting from $|\Phi\rangle$, is given by $\rho^{\mbox{\scriptsize gW\normalsize}}
=pP[|000\rangle]+(1-p)P[|\Phi\rangle]$, leading to 
$\rho^{\mbox{\scriptsize gW\normalsize}}_{12}=[p+(1-p)|a_0|^2] P[|00\rangle]+(1-p)P[|\psi\rangle]$, while 
$\rho^{\mbox{\scriptsize gW\normalsize}}_{13}$ is obtained by interchanging $a_0$ and $a_1$
in $\rho^{\mbox{\scriptsize gW\normalsize}}_{12}$.
In case of the PD channel, we define the states 
$|\tilde{\psi}\rangle=(1-p)(h_0a_0|001\rangle+h_1a_1|010\rangle+h_2a_2|100\rangle)$, and 
$|\tilde{\phi}\rangle=h_1a_1|01\rangle+h_2a_2|10\rangle$, so that
$h_ih_j =(1-p)^{-2}$ if $i=j$, and $1$ otherwise.
In terms of $|\tilde{\psi}\rangle$ and $|\tilde{\phi}\rangle$,  $\rho^{\mbox{\scriptsize gW\normalsize}}=P[|\tilde{\psi}\rangle]$, 
and $\rho^{\mbox{\scriptsize gW\normalsize}}_{12}=|a_0|^2P[|00\rangle]+(1-p)P[|\tilde{\phi}\rangle]$, respectively. 
Again, $\rho^{\mbox{\scriptsize gW\normalsize}}_{13}$ can be 
obtained from $\rho^{\mbox{\scriptsize gW\normalsize}}_{12}$ by interchanging $a_0$ and $a_1$. The form of 
$\rho^{\mbox{\scriptsize gW\normalsize}}$ in the case of the DP channel is given by 
$\rho^{\mbox{\scriptsize gW\normalsize}}
=\sum_{i=1}^{3}|a_{i-1}|^2\varrho^{\otimes(i-1)}\otimes\varrho^\prime\otimes\varrho^{\otimes(3-i)}
+(2q-1)^2\Big[\Big(a_0a_1^*\varsigma_1+a_0a_2^*\varsigma_2+a_1a_2^*\varsigma_3\Big)+h.c.\Big]$,
where $\varrho=qP[|0\rangle]+(1-q)P[|1\rangle]$, $\varrho^\prime=(1-q)P[|0\rangle]+qP[|1\rangle]$, with
$\varsigma_1=\varrho\otimes|0\rangle\langle1|\otimes|1\rangle\langle0|$,
$\varsigma_2=|0\rangle\langle1|\otimes\varrho\otimes|1\rangle\langle0|$,
$\varsigma_3=|0\rangle\langle1|\otimes|1\rangle\langle0|\otimes\varrho$.
The two-qubit reduced states, $\rho^{\mbox{\scriptsize gW\normalsize}}_{12}$ and $\rho^{\mbox{\scriptsize gW\normalsize}}_{13}$, 
can be obtained from the form of $\rho^{\mbox{\scriptsize gW\normalsize}}$ by 
tracing out qubit $3$ and $2$ respectively. As in the case of global noise, local density matrices, up to certain value of the 
noise parameter, remains quantum correlated.

In case of the global noise, negativity score, $\delta_\mathcal{N}^g$, is given by  
\begin{eqnarray}
 \delta_{\mathcal{N}}^g&=&\Big|\min\Big[0,s^g\Big]\Big|-\Big|\min\Big[0,s^g_{12}\Big]\Big|
 -\Big|\min\Big[0,s^g_{13}\Big]\Big|, 
\end{eqnarray}
with $s^g=\frac{p}{8}-(1-p)|a_2|\sqrt{1-|a_2|^2}$, and 
$s^g_{12}=\frac{1}{4}\Big[p+2(1-p)(|a_0|^2-\sqrt{|a_0|^4+4|a_1|^2|a_2|^2})\Big]$,
while for the PD channel, $\delta_{\mathcal{N}}^{pd}$ is obtained as  
\begin{eqnarray}
  \delta_{\mathcal{N}}^{pd}&=&s^{pd}-\frac{1}{2}\Big(s^{pd}_{12}+s^{pd}_{13}+|a_2|^2-1\Big),
\end{eqnarray}
where $s^{pd}=(1-p)^2|a_2|^2([1-|a_2|^2)^{\frac{1}{2}}$, $s^{pd}_{12}=[|a_0|^2+4|a_1|^2|a_2|^2(1-p)^4]^{\frac{1}{2}}$. 
In both the cases, $s^g_{13}$ and $s^{pd}_{13}$ are obtained from $s^{g}_{12}$ and $s^{pd}_{12}$, respectively, 
by interchanging $|a_0|$ and $|a_1|$. The expressions for negativity score, $\delta_{\mathcal{N}}^{ad}$, in the case of AD channel, 
is given by 
\begin{eqnarray}
 \delta_{\mathcal{N}}^{ad}&=&\frac{1}{2}\Big[(s^{ad}-p)-(s^{ad}_{12}-\tilde{p}_0)-(s^{ad}_{13}-\tilde{p}_1)\Big],
\end{eqnarray}
where $s^{ad}=\sqrt{p^2+4(1-|a_2|^2)|a_2|^2(1-p)^2}$, and $s^{ad}_{12}=\sqrt{\tilde{p}_0^2+4|a_1|^2|a_2|^2(1-p)^2}$,
with $\tilde{p}_j=p+(1-p)(\delta_{j0}|a_0|^2+\delta_{j1}|a_1|^2)$. 
Here also, the function $s^{ad}_{13}$ is obtained from $s^{ad}_{12}$
by interchanging $a_0$ and $a_1$. The expression for $\delta_\mathcal{N}^{dp}$, in the case of the DP channel, can also be obtained following 
the same procedure as in the cases of other three types of noise. However, the expression is rather involved, and to keep the text uncluttered, 
we choose not to include the expression.

\section{Computation of discord monogamy score for generalized W states: Numerical analysis}
\label{sec:disc_num_adv}

The computation of $\delta_{\mathcal{D}}$, in case of the gW states under noise,
requires more numerical resources than that in the case of the gGHZ states, since both 
$\mathcal{D}\big(\rho^{\mbox{\scriptsize gW\normalsize}}_{12}\big)$
and $\mathcal{D}\big(\rho^{\mbox{\scriptsize gW\normalsize}}_{13}\big)$ do not vanish for almost all  $p$. 
In the present case, $\delta_\mathcal{D}$ can be written as $\delta_\mathcal{D}=S-S(\rho^{\mbox{\scriptsize gW\normalsize}}_1)-S_c$,
where $S=S(\rho^{\mbox{\scriptsize gW\normalsize}}_{12})+S(\rho^{\mbox{\scriptsize gW\normalsize}}_{13})
-S(\rho^{\mbox{\scriptsize gW\normalsize}})$, and $S_c=S(\rho^{\mbox{\scriptsize gW\normalsize}}_{2|1})
+S(\rho^{\mbox{\scriptsize gW\normalsize}}_{3|1})-S(\rho^{\mbox{\scriptsize gW\normalsize}}_{23|1})$.
The determination of $\delta_\mathcal{D}$ for a single three-qubit state requires, in principle, three separate optimizations
for the terms in $S_c$.  
However, information acquired via numerical analysis using constrained optimization 
\cite{S_num-err_group} may result in considerable reduction of the computational complexity.

Let us first concentrate on the computation of $\mathcal{D}(\rho^{\mbox{\scriptsize gW\normalsize}}_{1:23})$ under four 
types of noise considered in this paper.
We perform extensive numerical search by Haar uniformly generating a set of  
$3\times10^6$ random three-qubit states of the form $\rho^\prime$ for each of the types of noise considered in this paper.
We find that for all such states, 
considering two sets of values of the real parameters, $(\theta,\phi)$, in projection measurements involved in 
$\mathcal{D}(\rho^{\mbox{\scriptsize gW\normalsize}}_{1:23})$, is enough. 
These sets are given by \textbf{(i)} $\theta=\pi/2,0\leq\phi<2\pi$, and 
\textbf{(ii)} $\theta=0,\pi,0\leq\phi<2\pi$,  which correspond to projection measurements on the $(x,y)$ plane, 
and along the $z$ axis of the Bloch sphere, respectively. Without any loss of generality, one can consider a 
projection measurement corresponding to the observable $\sigma_x$ in the former case, while a projection measurement corresponding to 
$\sigma_z$ in the latter. We refer to the set of states of the form $\rho^{\mbox{\scriptsize gW\normalsize}}$, for 
which measurement corresponding to $\sigma_x$, or $\sigma_z$ provides the optimal measurement, as the ``special'' set, 
denoted by $\mathcal{S}$. In the present case, the set $\mathcal{S}$ represents the set of all 
states of the form $\rho^{\mbox{\scriptsize gW\normalsize}}$, 
for each of the types of noise, according to our numerical analysis. 
The set of states for which the optimization occurs for $\sigma_x$,
is denoted by $\mathcal{S}_x$, while $\mathcal{S}_z$ represents the set of $\rho^{\mbox{\scriptsize gW\normalsize}}$ 
for which optimal measurement 
corresponds to $\sigma_z$. Note that $\mathcal{S}=\mathcal{S}_x\cup\mathcal{S}_z$, while 
$\mathcal{S}_x\cap\mathcal{S}_z=\Phi$, the null set.

\begin{table*}
 \begin{center}
 \begin{tabular}{|c|c|c|c|c|}
  \hline
  & \textbf{(I)} $\%$ of states $\in\mathcal{S}_x$ & \textbf{(II)} $\%$ of states $\in\mathcal{S}_z$ 
  & \textbf{(III)} $\%$ of states $\notin\mathcal{S}$ & \textbf{(IV)} Values of $\varepsilon_{max}$ \\
  \hline 
  \begin{tabular}{c}
  Noise-types \\
  \hline
  Global \\
  \hline 
  AD\\
  \hline 
  PD\\
  \hline
  DP\\  
  \end{tabular}
  & 
  \begin{tabular}{|c|c|}
  $\rho^{\mbox{\scriptsize gW\normalsize}}_{1:23}$ & $\rho^{\mbox{\scriptsize gW\normalsize}}_{12}$ \\
  \hline 
  $13.4\times10^{-2}$ & $76.632$ \\
  \hline 
  $99.937$ & $99.9027$ \\
  \hline 
   $9.73\times10^{-2}$ & $9.617$ \\
  \hline 
  $96.479$ & $95.735$ \\
  \end{tabular} 
  & 
  \begin{tabular}{|c|c|}
  $\rho^{\mbox{\scriptsize gW\normalsize}}_{1:23}$ & $\rho^{\mbox{\scriptsize gW\normalsize}}_{12}$ \\
  \hline 
  $99.866$ & $23.367$ \\
  \hline 
  $6.3\times10^{-2}$ & $9.73\times10^{-2}$ \\
  \hline 
  $99.9027$ & $90.373$ \\
  \hline 
  $3.521$ & $4.265$ \\
  \end{tabular}  
  & 
  \begin{tabular}{|c|c|}
  $\rho^{\mbox{\scriptsize gW\normalsize}}_{1:23}$ & $\rho^{\mbox{\scriptsize gW\normalsize}}_{12}$ \\
  \hline 
  $0$ & $1\times10^{-3}$ \\
  \hline 
  $0$ & $0$ \\
  \hline 
  $0$ & $1\times10^{-2}$ \\
  \hline 
  $0$ & $0$ \\
  \end{tabular}
  &
  \begin{tabular}{|c|c|}
  $\rho^{\mbox{\scriptsize gW\normalsize}}_{1:23}$ & $\rho^{\mbox{\scriptsize gW\normalsize}}_{12}$ \\
  \hline 
  $0$ & $1.23\times10^{-3}$ \\
  \hline 
  $0$ & $0$ \\
  \hline 
  $0$ & $2.75\times10^{-3}$ \\
  \hline 
  $0$ & $0$ \\
  \end{tabular}  
  \\
  \hline 
 \end{tabular}
 \end{center}
 \caption{Percentages of the states of the form $\rho^{\mbox{\scriptsize gW\normalsize}}$ and $\rho^{\mbox{\scriptsize gW\normalsize}}_{12}$, 
 belonging to the sets $\mathcal{S}_x$ and $\mathcal{S}_z$, are given in the columns \textbf{(I)} and \textbf{(II)} for 
 gW states subjected to different 
 noise models. The fraction of states of the form $\rho^{\mbox{\scriptsize gW\normalsize}}$ and 
 $\rho^{\mbox{\scriptsize gW\normalsize}}_{12}$, which do not belong to either of 
 $\mathcal{S}_x$ or $\mathcal{S}_z$, are given in the column \textbf{(III)}. 
 The upper bound of the absolute error, $\varepsilon_{max}$, is 
 given in column \textbf{(IV)} for different types of noise considered in this paper. In each column, the first sub-column corresponds to the 
 states of the form $\rho^{\mbox{\scriptsize gW\normalsize}}$, while the second is for $\rho^{\mbox{\scriptsize gW\normalsize}}_{12}$.}
 \label{tab2}
\end{table*}

The situation is a little different in the case of the two-qubit states $\rho^{\mbox{\scriptsize gW\normalsize}}_{12}$ and 
$\rho^{\mbox{\scriptsize gW\normalsize}}_{13}$, obtained from $\rho^{\mbox{\scriptsize gW\normalsize}}$. 
We generate $3\times10^6$ states Haar uniformly, which are of the form $\rho^{\mbox{\scriptsize gW\normalsize}}_{12}$, 
and we find that, like in the case of $\mathcal(\rho^{\mbox{\scriptsize gW\normalsize}}_{1:23})$, there exists, for each type of noise, 
a ``special'' set, $\mathcal{S}$, of states $\rho^{\mbox{\scriptsize gW\normalsize}}_{12}$, for which optimization occurs corresponding to either $\sigma_x$, 
or $\sigma_z$. 
However, in the case of global noise and PD channel, a small fraction of $\rho^{\mbox{\scriptsize gW\normalsize}}_{12}$ does not belong to 
$\mathcal{S}$, and the optimization of $\mathcal{D}(\rho^{\mbox{\scriptsize gW\normalsize}}_{12})$, for these states, occur for 
other values of $(\theta,\phi)$. 
Let the maximum absolute error, resulting from the assumption that all the three-qubit states of the 
form $\rho^{\mbox{\scriptsize gW\normalsize}}_{12}$ 
belong to $\mathcal{S}$, in the case of the global noise and PD channel, is $\varepsilon$. 
Our numerical analysis provides an upper bound of $\varepsilon$, denoted by $\varepsilon_{max}$, which is of the order of $10^{-3}$ in the 
case of both types of noise. 
Table \ref{tab2} displays our findings regarding 
the percentages of states of the form $\rho^{\mbox{\scriptsize gW\normalsize}}$ and $\rho^{\mbox{\scriptsize gW\normalsize}}_{12}$, 
that belong to the 
sets $\mathcal{S}_x$, $\mathcal{S}_z$, and do not belong to $\mathcal{S}$ for all the four types of noise. 
The last column (column \textbf{(IV)}) tabulates the values of $\varepsilon_{\max}$ in the 
relevant cases.
From now on, unless otherwise mentioned, we determine the values of $\delta_\mathcal{D}$ by computing quantum discord 
with the assumption that the states either belong to $\mathcal{S}_x$ or $\mathcal{S}_z$.

\section{Types of dynamics with generalized W states as input} 
\label{sec:type_dyne}

In case of the gW state subjected to 
global or local noise, the set of different types of dynamics that $\delta_\mathcal{N}$ and $\delta_\mathcal{D}$
undergo is far richer compared to that for gGHZ states. While only monotonic decay of $\delta_{\mathcal{N}}$ and $\delta_\mathcal{D}$ with 
increasing $p$ is found in the latter case, non-monotonic dynamics of monogamy scores emerges in the former.  
As an example, consider the set of gW states having a fixed value of $|a_2|$. 
The states in the set can be represented by the different allowed values of the absolute value of the free parameter, $a_0$.
Fig. \ref{gwnm} depicts the landscapes of $\delta_\mathcal{N}$, in the case of global 
noise, and of $\delta_\mathcal{D}$, in the case of the AD channel, as functions of 
$|a_0|$ and $p$, for $|a_2|=0.7$. The solid lines in the figures represent contours obtained by joining the points having 
a constant value of either $\delta_\mathcal{N}$, or $\delta_\mathcal{D}$. Note that the contours form closed curves, and  
from outside to inside, the lines represent increasing values of $\delta_\mathcal{N}$ and $\delta_\mathcal{D}$. 
The dashed lines in the plots represent the dynamics of $\delta_\mathcal{N}$, in the case of global noise, and 
$\delta_\mathcal{D}$, in the case of the AD channel, when the input gW state is taken with $|a_0|=0.5$. The behavior of 
the monogamy scores with increasing values of $p$ are non-monotonic, as clearly indicated from the values of 
$\delta_\mathcal{N}$ and $\delta_\mathcal{D}$, represented by different shades in Fig. \ref{gwnm}. 
An increase in the monogamy scores can be argued to be a signature of increase in quantumness. 
Although noise destroys quantum correlations, here we see the opposite by obtaining non-monotonicity of monogamy 
score with the increase of $p$.

\begin{figure}
 \includegraphics[scale=0.335]{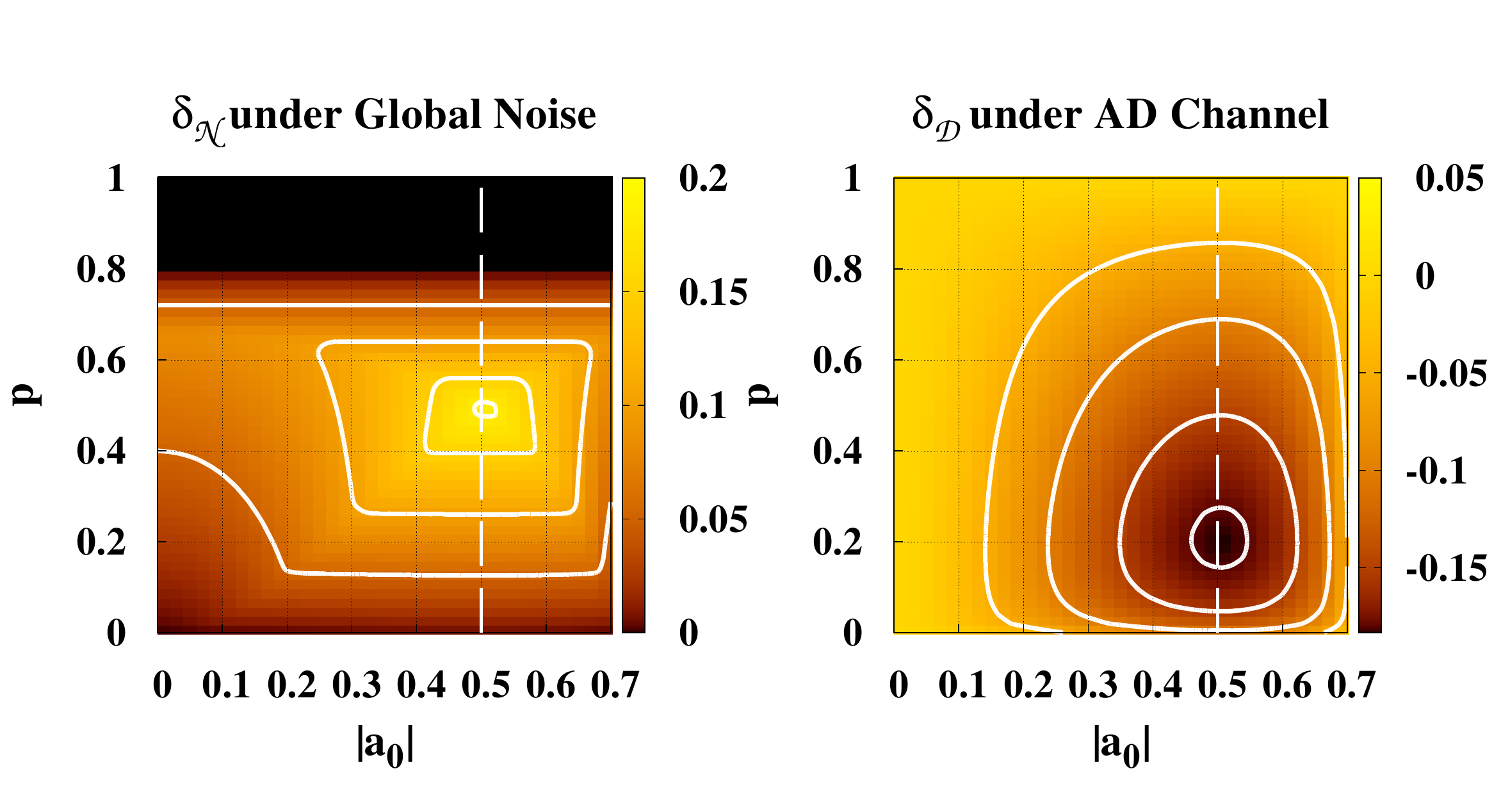}
 \caption{(Color online.) Variation of monogamy score corresponding to (a) $\delta_{\mathcal{N}}$ in the case of 
 global noise, and (b) $\delta_\mathcal{D}$ in the case of amplitude damping channel, as functions of the 
 state parameter $|a_0|$, and the noise parameter $p$, when gW state is subjected to noise. The value of $|a_2|$ is 
 fixed at $0.7$, while the value of $|a_1|$ is determined via normalization. 
 The dynamics of monogamy score along the dashed line at $|a_0|=0.5$ is non-monotonic in both the cases. 
 The solid lines represent the contours obtained by joining the points at which 
 $\delta_\mathcal{N}$, or $\delta_\mathcal{D}$ has a fixed value. From outside to inside, the closed contours correspond to 
 (a) $\delta_\mathcal{N}=0.05,0.10,0.15,0.18$, and (b) $\delta_\mathcal{D}=-0.05,-0.10,-0.15,-0.18$. 
 All the quantities plotted are dimensionless, except $\delta_\mathcal{D}$, which is in bits.}
 \label{gwnm}
\end{figure}

\noindent\textbf{Types of dynamics.}
Now we catalog four ``typical'' dynamics profiles observed for both $\delta_\mathcal{N}$ and $\delta_\mathcal{D}$ for global noise as 
well as for AD, PD, and DP local channels. 
\textbf{a.} In the first profile, $\delta_\mathcal{Q}(p=0)\geq0$, and $\delta_\mathcal{Q}(p)$ goes to 
zero non-monotonically as $p\rightarrow 1$. \textbf{b.} For the second one, 
$\delta_\mathcal{Q}(p)$ monotonically goes to zero when $p$ increases, with $\delta_\mathcal{Q}(p=0)\geq0$. 
\textbf{c.} In contrast to the first two profiles, $\delta_\mathcal{Q}(p=0)<0$ for the third profile. With an increase
of $p$, $\delta_\mathcal{Q}$ vanishes non-monotonically. \textbf{d.} Similar to the third profile, the 
fourth and the final profile starts with a non-monogamous scenario ($\delta_\mathcal{Q}(p=0)<0$). However, with increasing $p$, 
$\delta_\mathcal{Q}(p)$ goes to zero monotonically as $p\rightarrow 1$.  

Evidently, the frequencies of occurrence of the dynamics types \textbf{a}, \textbf{b}, \textbf{c}, and \textbf{d} 
must vary for different types of noise, and for different observables, viz., $\delta_{\mathcal{N}}$, and $\delta_\mathcal{D}$.   
To estimate these, we prepare a sample of $10^6$ Haar-uniformly generated gW states as input, which can be subjected to each of the 
types of noise, and study the dynamics profiles of the states. 
We find that at $p=0$, $75.195\%$ of the gW states are monogamous when 
negativity is considered, while only $13.333\%$ of them are monogamous 
with respect to quantum discord.    
When the value of $p$ is increased, the four types of dynamics are found to occur 
with different frequencies in the case of the global noise and the local channels (see Table. \ref{tab3}).  
Note that for $\delta_\mathcal{N}$, type-\textbf{a} is more frequent in the case of 
global noise as well as for the PD and DP channels, while type-\textbf{b} occurs mostly in the case of the AD channel.
The frequency of occurrence of \textbf{d} is much less compared to that of \textbf{a}, \textbf{b}, and \textbf{c} 
for the negativity monogamy score. 
Among all the noisy channels, the non-monotonic decay of $\delta_\mathcal{N}$ occurs close to
$100\%$ of times when global noise acts on the gW state, irrespective of the sign of $\delta_\mathcal{N}$ at $p=0$.
On the other hand, in the case of $\delta_\mathcal{D}$, frequency of 
occurrence of \textbf{c} and \textbf{d} is high in the cases of global noise and the AD channel, while the same is moderate in the case 
of the PD and the DP channels.

\begin{table*}
 \begin{center}
 \begin{tabular}{|c|c|c|c|c|}
  \hline
 \cellcolor{blue!25} & \textbf{a} & \textbf{b} & \textbf{c} & \textbf{d} \\
  \hline 
  \begin{tabular}{c}
  Noise-types \\
  \hline
  Global \\
  \hline 
  AD\\
  \hline 
  PD\\
  \hline
  DP\\  
  \end{tabular}
  & 
  \begin{tabular}{c|c} % a
  $\delta_\mathcal{N}$ & $\delta_\mathcal{D}$ \\
  \hline 
  $75.009$ &  $13.330$\\
  \hline 
  $17.532$ & $13.321$ \\
  \hline 
  $56.145$ & $13.323$ \\
  \hline 
   $53.531$ & $8.708$ \\
  \end{tabular} 
  & 
  \begin{tabular}{c|c} % b
  $\delta_\mathcal{N}$ & $\delta_\mathcal{D}$ \\
  \hline 
  $0.186$ & $0.003$ \\
  \hline 
  $57.663$ & $0.012$ \\
  \hline 
  $19.050$ & $0.010$ \\
  \hline 
  $21.664$ & $4.625$ \\
  \end{tabular}  
  & 
  \begin{tabular}{c|c} % c
  $\delta_\mathcal{N}$ & $\delta_\mathcal{D}$ \\
  \hline 
  $24.805$ & $83.536$ \\
  \hline 
  $23.964$ & $82.902$ \\
  \hline 
  $24.784$ & $53.272$ \\
  \hline 
   $24.802$ & $68.710$  \\
  \end{tabular}
  &
  \begin{tabular}{c|c} % d
  $\delta_\mathcal{N}$ & $\delta_\mathcal{D}$ \\
  \hline 
  $0.000$ & $0.131$ \\
  \hline 
  $0.841$ & $3.765$ \\
  \hline 
  $0.021$ & $33.395$ \\
  \hline 
  $0.003$ & $17.957$ \\
  \end{tabular}  
  \\
  \hline 
 \end{tabular}
 \end{center}
 \caption{The percentage of gW states exhibiting \textbf{a}, \textbf{b}, \textbf{c}, and \textbf{d}-type dynamics for 
 $\delta_\mathcal{N}$ and $\delta_\mathcal{D}$ under 
 the application of different types of noise.}
 \label{tab3}
\end{table*}

\section{Dynamics terminal for generalized W states}  
\label{sec:dyn_term}

To check whether the dynamics terminal, $\langle p_t\rangle$, can distinguish between different types of noise, one has to 
determine the value of $\langle p_t\rangle$ for different types of noise with gW states as input states, which, in turn, requires the
determination of $P(p_t)$. 
In order to determine $P(p_t)$, we Haar uniformly generate $10^6$ gW states 
for each of the four kinds of noise, and study their dynamics profiles to determine $\langle p_t\rangle$.
The variations of $P(p_t)$ against $p_t$ are given in Fig. \ref{pdn}. It is clear from the figure that 
the maximum possible value of $p_t$ is considerably different in the case of $\delta_\mathcal{N}$ and $\delta_\mathcal{D}$,
when the type of noise is fixed. The values of $\langle p_t\rangle$ corresponding to $\delta_\mathcal{N}$ and 
$\delta_\mathcal{D}$, for global noise, AD channel, PD channel, and DP channel are 
given in Table \ref{tab4}.  
Note that the dynamics terminal corresponding to $\delta_\mathcal{D}$ is higher 
than that corresponding to $\delta_\mathcal{N}$ in the case of the global noise, AD channel,  and the DP channel, while the trend is 
reversed in the case of the PD channel. 
% Note also that the behavior of $\langle p_t\rangle$ corresponding to $\delta_\mathcal{N}$, in the results obtained 
% previously in Sec. \ref{sec:mod_noise} in the case of the PD channel, is in good agreement with the value of $\langle p_t\rangle$ 
% corresponding to $\delta_\mathcal{N}$, which is maximum for the PD channel, and minimum for the DP channel.
% It is interesting to note here that there can not be any bipartite state for which $\mathcal{N}>0$ while $\mathcal{D}=0$. However, such a 
% situation is possible in the case of $\delta_\mathcal{N}$ and $\delta_\mathcal{D}$. We shall discuss this issue in Sec. \ref{sec:sd_dms}
% in detail.

\begin{figure*}
 \begin{center}
 \includegraphics[width=\textwidth]{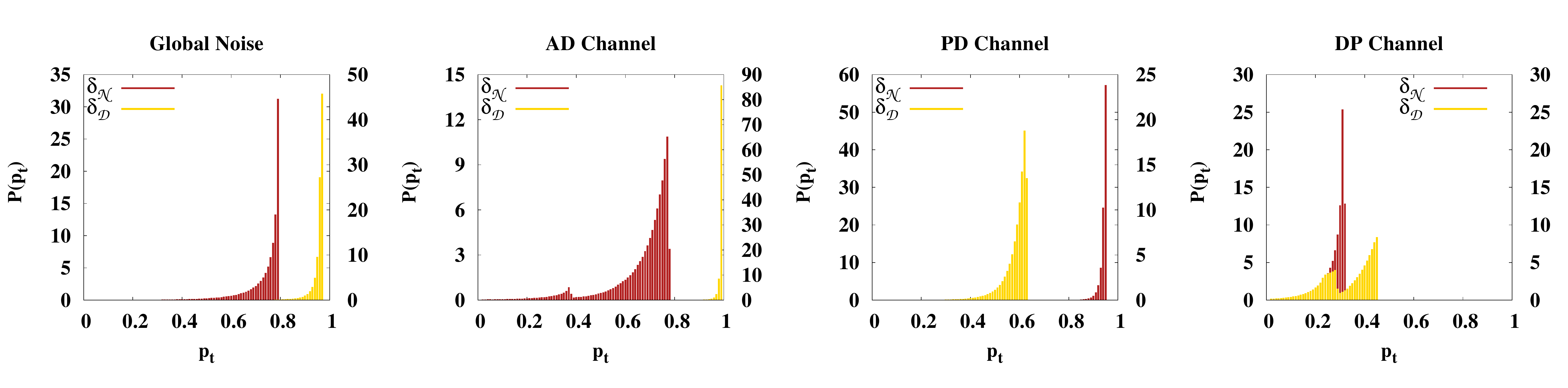}
 \end{center}
 \caption{(Color online.) Variations of the normalized probability density function, $P(p_t)$, against the dynamics terminal,
 $p_t$, for $\delta_{\mathcal{N}}$ and $\delta_\mathcal{D}$, when gW states are subjected to different types of noise. All 
 quantities plotted are dimensionless.}
 \label{pdn}
\end{figure*}

\begin{figure*}
 \begin{center}
 \includegraphics[width=\textwidth]{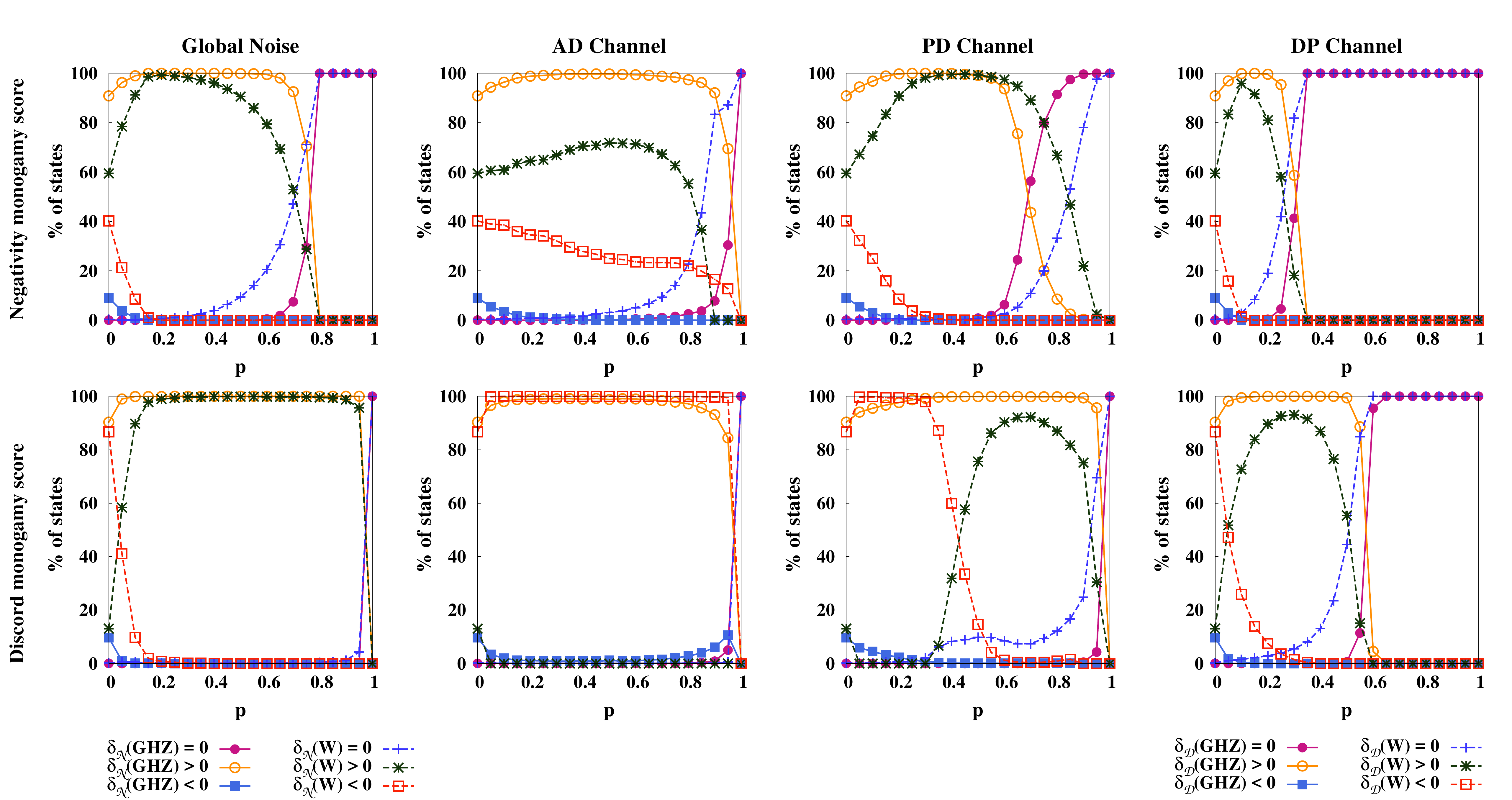} 
 \end{center}
 \caption{(Color online.)
 Variations of the percentages of arbitrary three-qubit pure states chosen from GHZ class (continuous lines) 
 and W class (broken lines), 
 for which monogamy scores corresponding to negativity and quantum discord are strictly greater than, equal to, and strictly less than zero
 (denoted by different types of points), 
 with the noise parameter, $p$. All quantities plotted are dimensionless.}
 \label{ghzwc}
\end{figure*}

\section{Dynamics of monogamy score in arbitrary tripartite pure states}

Hitherto, we have investigated gGHZ and gW states, for which the effects of various noisy channels on monogamy scores can be addressed 
analytically up to certain extent. To complete the investigation for three-qubit states as input, we now consider the two
mutually exclusive and exhaustive classes of three-qubit  
states, viz., the GHZ class and the W class \cite{S_dvc}. 
These two classes, inequivalent under stochastic local operations and classical 
communication (SLOCC), together span the entire set of three-qubit pure states \cite{S_dvc}.
An arbitrary three-qubit pure state from the GHZ class, up to local unitary operations, can be parametrized as 
$|\psi_{GHZ}\rangle=\sqrt{K}(c_{\delta}|000\rangle+s_{\delta}e^{i\varphi}|\varphi_{\alpha}\rangle|\varphi_{\beta}
\rangle|\varphi_{\gamma}\rangle)$, where $|\varphi_k\rangle=c_{k}|0\rangle+s_{k}|1\rangle$ with 
$c_k=\cos k$, $s_k=\sin k$, $k=\alpha, \beta, \gamma$, and 
$K=(1+2c_{\delta}s_{\delta}c_{\alpha}c_{\beta}c_{\gamma}c_{\varphi})^{-1}\in(\frac{1}{2},\infty)$
is the normalization factor. Here, the ranges for the five real parameters are $\delta \in (0,\pi/4], \alpha,\beta,\gamma \in 
(0,\pi/2]$ and $\varphi \in [0,2\pi)$. On the other hand, a three-qubit pure state from the W class, up to local unitaries, 
can be written in terms of three real parameters as
$|\psi_{W}\rangle=\sqrt{a}|001\rangle+\sqrt{b}|010\rangle+\sqrt{c}|100\rangle+\sqrt{1-(a+b+c)}|000\rangle$,
where $a,b,c\geq0$. Due to higher number of state parameters in arbitrary three-qubit pure states chosen from these classes, 
determining compact forms for $\delta_{\cal N}$ as well as $\delta_{\cal D}$ is difficult. Also, the constrained optimization
is not applicable due to the high absolute error in the value of quantum discord. Therefore, we  
employ exact numerical optimization technique to compute quantum discord in discord monogamy scores 
of these states. We Haar-uniformly generate $10^4$ states from each of the two classes -- the GHZ class and the W class -- 
for a chosen value of the noise parameter, $p$, when a specific type of noise is applied to it. 
We then determine the percentage of states for which negativity and discord monogamy scores are greater than, equal to, and less than zero, 
and study the variation of these percentages with varying noise parameter.

The variation of the percentages of three-qubit pure states from the GHZ and W classes, 
for which $\delta_\mathcal{N}$ and $\delta_\mathcal{D}$ are $>$,
$=$, and $<0$, against $p$ is given in Fig. \ref{ghzwc}. 
The percentages vary non-monotonically with varying noise parameter, and the percentage of states for which the monogamy scores corresponding 
to negativity and quantum discord are equal to zero, for both classes of states, tend to become $100\%$ with increasing $p$, as expected. 
For both the classes, this trend is considerably slower in the case of global noise, AD channel, and PD channel, in comparison to 
that for the DP channel. The patterns in the W class states are similar 
to those in the case of gW states, except for discord monogamy score under  
PD channel. While no gW states have a strictly positive $\delta_\mathcal{D}$ for higher values of $p$, in the case of W class states, the 
corresponding fraction increases with increasing $p$, reaches a maximum value at moderately high $p$, and then, as expected, 
decreases to zero as $p\rightarrow1$.

\begin{figure}
 \begin{center}
 \includegraphics[scale=0.4]{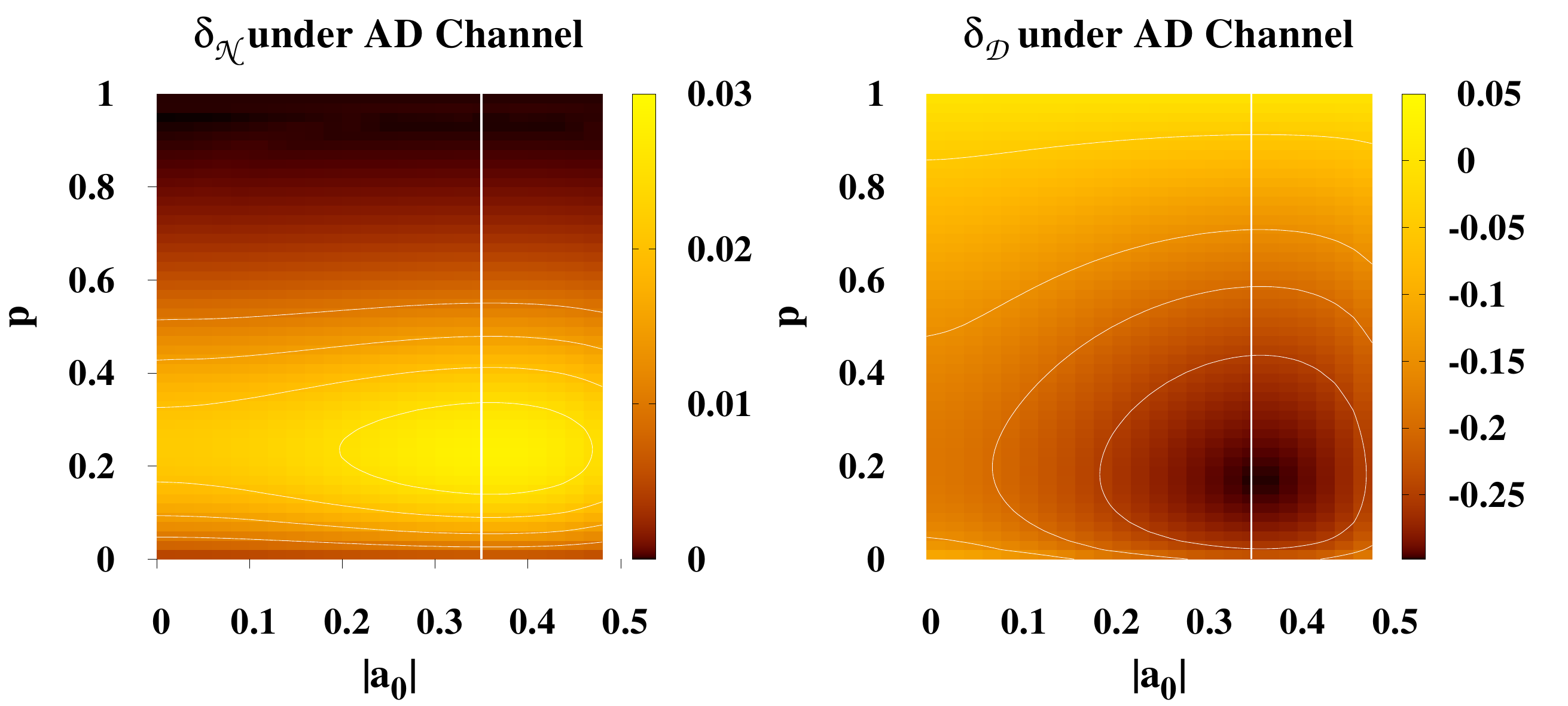}
 \end{center}
 \caption{(Color online.) Variation of monogamy scores corresponding to (a) $\delta_{\mathcal{N}}$ 
 and (b) $\delta_\mathcal{D}$ in the case of the amplitude damping channel, as functions of the 
 state parameter $|a_0|$, and the noise parameter $p$, when the four-qubit gW state is subjected to the noise. 
 The value of $|a_2|$  and $|a_3|$ are 
 fixed at $0.5$ and $0.7$, respectively, while the value of $|a_1|$ is determined via normalization. 
 The dynamics of monogamy scores along the thick solid lines at $|a_0|=0.35$ are non-monotonic in both the cases. 
 The thin dashed lines represent the contours obtained by joining the points at which 
 $\delta_\mathcal{N}$, or $\delta_\mathcal{D}$ have a fixed value. From outside to inside, the closed contours correspond to 
 (a) $\delta_\mathcal{N}=0.01,0.015,0.02,0.025$, and (b) $\delta_\mathcal{D}=-0.05,-0.15,-0.2,-0.25$. 
 All the quantities plotted are dimensionless, except $\delta_\mathcal{D}$, which is in bits.}
 \label{gwnm4}
\end{figure}

\section{Dynamics of monogamy score for higher number of parties}

The question that naturally arises from the results reported in the preceeding sections is 
whether the dynamical features found in the case of three-qubit systems is generic for quantum states with higher number of qubits. 
As we have already mentioned, the monogamy score of gGHZ state decreases monotonically with  the increase of $p$ for an 
arbitrary number of parties.  
And, the non-monotonic behaviour of the monogamy scores corresponding to both negativity and quantum discord,
in the case of the three-qubit generalized W state under, for example, the AD channel, is found in the case of higher number of qubits. 
This can be clearly seen in Fig. \ref{gwnm4}, 
where the variations of $\delta_{\mathcal{N}}$ and $\delta_{\mathcal{D}}$, in the case of the four-qubit generalized W state 
$|\Phi_4\rangle=a_0|0001\rangle+a_1|0010\rangle+a_2|0100\rangle+a_3|1000\rangle$, is depicted as functions of the noise parameter $p$, 
and the state parameter, $|a_0|$. We fix $|a_3|=0.7$ and $|a_2|=0.5$ for demonstration, where $|a_1|$ is fixed by normalization. 
We also find that similar to the three-qubit scenario, the depolarizing channel destroys the monogamy scores corresponding to both 
negativity and quantum discord more rapidly compared to other channels, for four qubits. We therefore expect that
the broad qualitative features  of the results
found in the case of three qubits remain unaltered when the number of qubits is increased.

\end{document}